\documentclass{jfm}
\usepackage{epsfig,graphics,graphicx,amssymb,amsmath,natbib,upmath,multirow}
\usepackage[usenames,dvipsnames]{color}
\usepackage[sfdefault=cmbr]{isomath}
\usepackage{chngcntr}
\counterwithout{table}{section}

\usepackage{color, colortbl,booktabs}
\usepackage[normalem]{ulem}
\DeclareMathOperator{\arccot}{arccot}
\def\t{\tensorsym}

\def\b{\bm}\def\u{\underline}

\def\u{\gamma}

\definecolor{light-gray}{gray}{0.88}

\definecolor{DarkBlue}{rgb}{0,0,.8}
\newcommand\tcb{\textcolor{black}}

\include{help}

\title{Sedimentation of spheroidal bodies near walls in viscous fluids: glancing, reversing, tumbling, and sliding}
\shorttitle{Sedimentation of spheroidal bodies near walls in viscous fluids}

\author[W. H.~Mitchell and S.~E.~Spagnolie]%
{William H. Mitchell and Saverio~E.~Spagnolie\thanks{Email address for correspondence: spagnolie@math.wisc.edu}}

\affiliation{Department of Mathematics, University of Wisconsin--Madison, 480 Lincoln Drive, Madison, \\WI 53706, USA}

\date{\today}

\pubyear{2015}
\volume{??}
\pagerange{??--??}
\date{\today}
\usepackage{graphicx,color,bm}
\graphicspath{{converted_graphics/}}
\usepackage{graphicx}
\begin{document}

\maketitle

\begin{abstract}
The sedimentation of a rigid particle near a wall in a viscous fluid has been studied numerically by many authors, but analytical solutions have been derived only for special cases such as the motion of spherical particles. In this paper the method of images is used to derive simple ordinary differential equations describing the sedimentation of arbitrarily oriented prolate and oblate spheroids at zero Reynolds number near a vertical or inclined plane wall. The differential equations may be solved analytically in many situations, and full trajectories are predicted which compare favorably with complete numerical simulations. The simulations are performed using a novel double layer boundary integral formulation, a method of stresslet images. The conditions under which the glancing and reversing trajectories, first observed by \cite{rhlt77}, occur are studied for bodies of arbitrary aspect ratio. Several additional trajectories are also described: a periodic tumbling trajectory for nearly spherical bodies, a linearly stable sliding trajectory which appears when the wall is slightly inclined, and three-dimensional \tcb{glancing, reversing, and wobbling.}
\end{abstract}

\section{Introduction}
The sedimentation of bodies in viscous fluids is important in many natural settings and industrial processes, from paper manufacturing \citep{sj58} to blood circulation \citep{caro2012} to the settling of contaminant particles through oil in internal combustion engines \citep{Guazzelli06}. \tcb{The scientific study of these viscous sedimentation processes encompasses both analytical and numerical treatments dating back to the work of \cite{Stokes1851} on the flow past a sphere in an unbounded fluid.} A number of exact solutions have since been derived to describe the dynamics of sedimenting bodies of simple shape and symmetric orientation.  A generalization known as the Fax\'en law gives formulas for the force and torque on a sphere placed in an arbitrary background flow \citep{Faxen1922,Faxen1924}. \cite{sj1926} considered two sedimenting spheres of equal density and radius, with one placed directly above the other, and showed that the settling speed is increased by their interaction through the fluid.  Exact series solutions for two arbitrarily oriented identical spheres were then derived \citep{gcb66}. Single non-spherical bodies were also treated classically by \cite{Oberbeck1876}, \cite{Edwardes1892}, \cite{Jeffery22}, and \cite{Lamb1932} who found the force and torque on a triaxial ellipsoid in a linear flow field in terms of ellipsoidal harmonics.  Later, \cite{cw75,cw76} gave a simpler solution of the same problem using the singularity method, in which \tcb{fundamental} solutions of the Stokes equations are placed internal to the body surface with coefficients selected so as to satisfy the no-slip boundary condition.  
For particle-wall interactions, an exact solution for a sphere translating and rotating near a plane wall was obtained by \cite{Oneill64}, \cite{Brenner1961} and \cite{gcb67}, and in a shear flow by \cite{gcb67b}.  The study of other body types generally requires methods of approximation such as exploiting particle slenderness, as in the various slender body theories \citep{Batchelor70,Cox70, tillett70,Lighthill76,kr76,Johnson80, Blake2010}, or weak particle flexibility \citep{lmss13}. 

A widely employed strategy for incorporating the hydrodynamic effect of a plane wall is the method of reflections, an iterative solution procedure where boundary conditions are alternately enforced on the particle and the wall, leading to asymptotically valid representations of fluid forces and particle velocities. Problems involving the extreme cases of spheres and rods have been more frequently investigated than spheroids of intermediate eccentricity; a prominent exception \tcb{is} work by \cite{Wakiya59} wherein the mobility of a general ellipsoid near a wall is approximated using reflections of Lamb's general solution \citep{Lamb1932}.  Various investigations using slender bodies include that of \cite{rhlt77} who derived an asymptotic expression for the rotation of a slender cylinder sedimenting near a plane wall, \tcb{ignoring end effects}; they observed two types of trajectories which they \tcb{termed} \emph{glancing} and \emph{reversing}.  A related result using matched asymptotic expansions is due to \cite{kbp75} who gave an analytical solution of the mobility problem for a slender rod near a wall.  Later, \cite{bl88} used resistive force theory to find simplified integral equations for the rigid motion of a (possibly non-straight) slender body near a wall, including the special case of a slender prolate body. \cite{yl83} studied the more general problem of motion near an interface between fluids of two different viscosities, giving asymptotically valid ordinary differential equations describing the trajectories of slender rods.

Many generalizations of the problem of a sphere moving near a wall in a viscous fluid have been investigated. A series of papers (\cite{bms91}, \cite{cjkw00}, and \cite{sb07}) addressed the problem of constructing the (positive-definite) grand mobility tensor for many spherical particles above a plane wall. The \tcb{extensive} literature on the subject includes \tcb{treatments of} the electrophoresis of a single charged nonconducting sphere near a wall \citep{ka85}, a deformable droplet moving between two parallel plates \citep{sh88}, and a colloidal sphere translating perpendicularly between two parallel plane walls \citep{kw08}. Similar efforts have been extended to the study of the trajectories of swimming microorganisms near surfaces (see \cite{sl12} and references provided therein). 

Particle dynamics in the presence of surfaces have also been \tcb{studied} numerically. \cite{hg89,hg94} computed the solution to the resistance problem for a spheroid of arbitrary aspect ratio near a plane wall \tcb{using a combined single- and double-layer representation of the flow. In addition to confirming the glancing and reversing scenarios found by \cite{rhlt77}, these authors studied the sedimentation problem for inclined walls and observed trajectories in which the particle escapes from the wall as well as a stable steady solution \tcb{where the particle translates parallel to the wall without rotation}. \cite{hhj98} and \cite{smh06} considered the sedimentation of a prolate body in a circular or rectangular cylinder at finite Reynolds number, \cite{Pozrikidis07} considered a sphere near the interface of two fluids of varying viscosity, and \cite{kutteh2010} \tcb{treated flows containing} several non-spherical particles by modeling irregular particles as rigid ensembles of spheres. Boundary integral methods are commonly used to solve particle-wall interaction problems numerically (see \cite{Pozrikidis92}), but novel numerical methods have also been developed, including the method of regularized Stokeslets with images by \cite{adebc08}}.

In this paper we consider the problem of sedimenting prolate and oblate spheroids of arbitrary aspect ratio in a wall-bounded Stokes flow. We present a unified description of the particle dynamics, combining and generalizing the trajectories observed by \cite{rhlt77} and \cite{hg94}, along with \tcb{several novel trajectory types}. Using the method of images, we derive asymptotically valid ordinary differential equations to describe the body dynamics, \tcb{accurate up to $O(h^{-4})$ in the translational velocity and $O(h^{-5})$ in the rotational velocity where $h$ is the distance from the particle centroid to the wall}. The resulting system is further reduced to yield analytical solutions for the complete particle trajectory in many cases, and the predictions are found to agree very well with full numerical simulations. These numerical simulations \tcb{are carried out using} a novel double layer boundary integral formulation \tcb{which we call the} method of stresslet images. We describe \tcb{various} types of trajectories that can arise during sedimentation near a wall, from glancing and reversing to periodic tumbling orbits \tcb{as well as the sliding trajectory which can arise if the wall is tilted relative to gravity.} We also generalize previously published work by treating arbitrary particle orientations instead of laterally symmetric configurations. This leads to fully three-dimensional dynamics, which may result in periodic wobbling, \tcb{three-dimensional} glancing, and \tcb{three-dimensional} reversing trajectories.

The paper is organized as follows. In \S\ref{sec:EqOfMotion} we describe the geometry of the problem and the equations of motion and we present the method of stresslet images. In \S\ref{sec:NumTraj} we conduct a numerical survey of the particle dynamics and present a qualitative overview of the zoology of trajectory types. In \S\ref{sec:moi} we apply the method of images to reduce the sedimentation problem for arbitrarily oriented spheroids to a two- or three- dimensional system of ordinary differential equations. In \S\ref{sec:SymbTraj}, we analyze this system of equations and provide closed-form results describing particle trajectories in many special cases; for example, a simple inequality indicates whether or not a particle of a given shape and initial data will escape from the wall. We conclude with a discussion of applications and possible directions for future work in \S\ref{sec:Discussion}.  

\section{Equations of motion and numerical method}
\label{sec:EqOfMotion}
\begin{figure}
\centering
 \includegraphics[width=.45\textwidth]{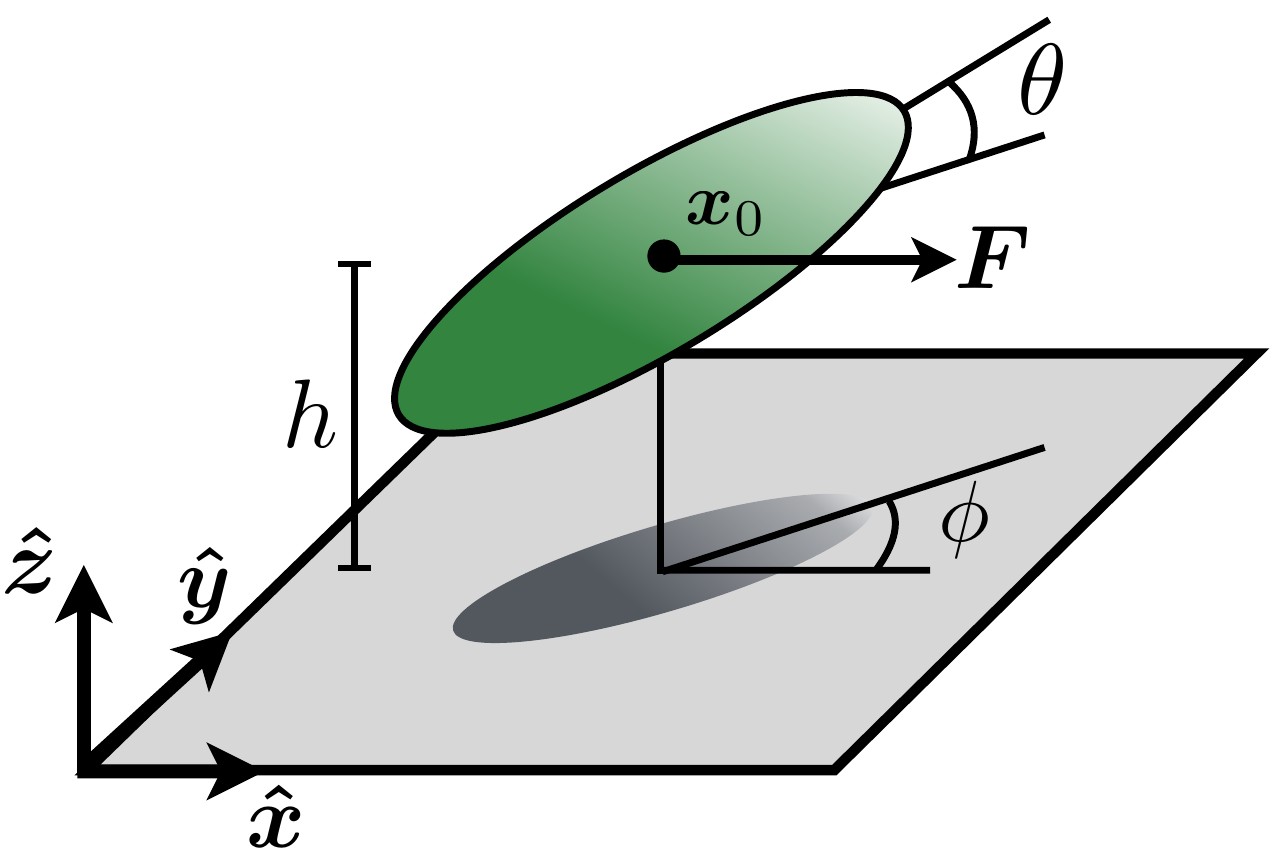}
\caption{Schematic of a prolate spheroid near an infinite plane wall located along the $x-y$ plane. The body, with centroid a distance $h$ from the wall, is rotated through an angle $\phi$ about the $\b{\hat{z}}$ axis and is pitched at an angle $\theta$ about its lateral axis. The dimensionless force due to gravity acts at an angle $\beta$ relative to the wall, $\b{F}=\cos\beta\b{\hat{x}}-\sin\beta\b{\hat{z}}$, with $\beta$ set to zero in the illustration above.}
\label{Schematic}
\end{figure}

Consider a spheroid \tcb{of uniform density} sedimenting through a viscous fluid near an infinite plane wall located along the $xy$-plane. The body surface, denoted by $S^*$, is described by
\begin{equation}
S^* = \left\{
a(x,y,h)^T+a\t{R}_\phi\cdot \t{R}_\theta\cdot(X,Y,Z)^T: X^2+(a/b)^2\,Y^2+(a/c)^2\,Z^2=1\bm \right\},
\label{eq:Geometry}
\end{equation}
where $a$ is a length, $a\b{x}_0=a(x,y,h)$ is the position of the body centroid, and 
\begin{gather}
\t{R}_\theta=\begin{pmatrix}\cos\theta&0&-\sin\theta\\0&1&0\\\sin\theta&0&\cos\theta\end{pmatrix}, \ \ \t{R}_\phi=\begin{pmatrix}\cos\phi&-\sin\phi&0\\\sin\phi&\cos\phi&0\\0&0&1\end{pmatrix},
\end{gather}
are rotation operators (see Fig.~\ref{Schematic}), with $\theta\in(-\pi/2,\pi/2]$ and $\phi=[0,2\pi)$. The semi-axis lengths satisfy $a>b=c$ for prolate spheroids and $a=b>c$ for oblate spheroids. The body eccentricity is given by $e = \sqrt{1-c^2/a^2}\in[0,1]$ in both cases, and the vectors $\b{x}=(x,y,h)^T$ and $\b{X}=(X,Y,Z)^T$ in Eq.~\eqref{eq:Geometry} are dimensionless. The body is subject to a gravitational force $\b{F}^*=\Delta \rho\,g V \b{\hat{x}}$, where $\Delta \rho$ is the density difference between the body and the fluid, $g$ is gravitational acceleration, and $V$ is the body volume; the study of particle sedimentation near an inclined wall is achieved by considering a gravitational force at an angle $\beta$ relative to $\b{\hat{x}}$. In response, the body moves with translational velocity $\b{U}^*$ and rotational velocity $\b{\Omega}^*$ which depend on the particle position and orientation. The system is made dimensionless by scaling lengths upon $a$ and defining the dimensionless translational and rotational velocities $\b{U}=(6\pi\mu a)(\Delta \rho\,g V)^{-1}\b{U}^*$ and $\bm{\Omega}=(6\pi\mu a^2)(\Delta \rho\,g V)^{-1}\bm{\Omega}^*$. In the case of a spherical particle in an unbounded fluid this results in a dimensionless body of radius and sedimentation speed both \tcb{equal to} unity, the well known Stokes drag law \citep{Stokes1851}. The surface $S$ denotes the dimensionless body surface (the surface $S^*$ scaled by the length $a$). 

In the theoretical limit of zero Reynolds number flow, \tcb{appropriate for modeling the flow generated by small or slow moving particles, or larger particles in highly viscous fluids}, the equations of fluid motion are the Stokes equations, 
\begin{gather}\label{eq: Stokes}
-\nabla p^*+\mu \nabla^2\b{u}^*=\b{0},\ \ \ \nabla \cdot \b{u}^*=0,
\end{gather}
where $p^*$ is the pressure, $\b{u}^*$ is the fluid velocity, and $\mu$ is the viscosity (see \cite{Batchelor00}). The dimensionless fluid velocity is also scaled as above, $\b{u}=(6\pi\mu a)(\Delta \rho\,g V)^{-1}\b{u}^*$, and is assumed to satisfy no-slip boundary conditions on the particle surface, $\b{u}(\b{x}\in S)=\b{U+\Omega}\times(\b{x-x}_0)$, and on the wall, $\b{u}(z=0)=\b{0}$. The fluid velocity is assumed to decay to zero as the distance from the body tends towards infinity. In the inertialess limit, the integrated fluid stress on the particle surface must balance the external force due to gravity, and there must be zero net fluid torque on the body. These six conditions, along with Eqs.~\eqref{eq: Stokes}, close the system of equations for the fluid velocity and pressure and the particle's instantaneous translational and rotational \tcb{velocities}, $\b{U}$ and $\b{\Omega}$. The position and orientation of the body at any time are described by the state variable $\b{\Phi}=(x,y,h,\theta,\phi)$. Since time does not appear explicitly in the Stokes equations, any solution of the mobility problem for general $\b{\Phi}$ yields an autonomous system of ordinary differential equations, $\dot{\b{\Phi}}= \b{\mathcal{F}}(h,\theta,\phi)$, describing the trajectory of the particle sedimenting under the influence of a constant gravitational force.




\subsection{Fundamental singularities and image systems}\label{sec:Singularities}

The linearity of the Stokes equations opens the door to numerous analytical and numerical approaches to solving fluid-body interaction problems that rely on fundamental singularities, or Green's functions. In one particularly useful and clean approach to solving such problems, the singularities are placed internal to an immersed body and their strengths are chosen so as to match the boundary conditions on the surface \citep{cw75}. For instance, the Stokes flow around a no-slip spherical boundary in an unbounded flow may be represented as a linear combination of a Stokeslet singularity,
\begin{gather}\label{eq: Stokeslet}
\t{G}(\b{x},\b{x}_0) = \frac{\t{I}}{|\b{x}-\b{x}_0|} + \frac{(\b{x}-\b{x}_0)(\b{x}-\b{x}_0)^T}{|\b{x}-\b{x}_0|^3},
\end{gather}
with $\t{I}$ the identity operator, and a potential source dipole (see \cite{kk91}). 

The effect of a wall on the trajectory of a moving body can be studied using image singularity systems \citep{Blake71,bc74}. Image systems for Stokeslets of varying orientation relative to a no-slip wall were presented by \cite{Blake71}. As an example, consider an $x$-directed Stokeslet $\b{G}_x(\b{x},\b{x}_0)=\t{G}(\b{x},\b{x}_0)\cdot \b{\hat{x}}$ located in the fluid at a point $\b{x}_0=(0,0,h)$. The image system cancels the fluid velocity on the surface $z=0$ when placed at the image point $\b{x}^*=(0,0,-h)$, and is given by
\begin{equation}\label{eq: Gstar}
\b{G}_x^*(\b{x},\b{x}^*)=-\b{G}_x(\b{x},\b{x}^*) - 2h\frac{\partial}{\partial x}\b{G}_z(\b{x},\b{x}^*)+2h^2\frac{\partial}{\partial x}\b{U}_P(\b{x},\b{x}^*),
\end{equation}
where $\b{G}_z=\t{G}\cdot \b{\hat{z}}$ is a $z$-directed Stokeslet and 
\begin{gather}
\b{U}_P(\b{x},\b{x}_0) = \frac{\b{x}-\b{x}_0}{|\b{x}-\b{x}_0|^3}
\end{gather}
is a potential flow point source. Similarly, 
\begin{equation}\label{eq: Gstarz}
\b{G}_z^*(\b{x},\b{x}^*)=-\b{G}_z(\b{x},\b{x}^*) + 2h\frac{\partial}{\partial z}\b{G}_z(\b{x},\b{x}^*)-2h^2\frac{\partial}{\partial z}\b{U}_P(\b{x},\b{x}^*).
\end{equation}
Image systems for derivatives of the Stokeslet may be determined by careful manipulation of the image systems of \cite{Blake71}, though some care must be taken as the image system of the derivative is not in general the derivative of the image system (see \cite{kk91} and \cite{sl12}). Note for instance that the coefficients in Eq.~\eqref{eq: Gstar} are $h$-dependent. Since $\nabla_{\b{x}}\t{G}(\b{x},\b{x}_0)=-\nabla_{\b{x}_0}\t{G}(\b{x},\b{x}_0)$, we may use Blake's result to construct the image system for a difference quotient.  As an example, the image system of $\partial^2\, \b{G}_x/\partial x\partial z$ is given by
\begin{multline}
\label{eq:Sx_xzimage}
\left[\frac{\partial^2}{\partial x\partial z}\b{G}_x\right]^*(\b{x},\b{x}^*)=-\frac{\partial^2}{\partial x\partial z}\b{G}_x(\b{x},\b{x}^*)
-2h\frac{\partial^3}{\partial x^2\partial z}\b{G}_z (\b{x},\b{x}^*)
\\+2\frac{\partial^2}{\partial x^2}\b{G}_z(\b{x},\b{x}^*)
+2h^2\frac{\partial^3}{\partial x^2\partial z}\b{U}_P(\b{x},\b{x}^*)
-4h\frac{\partial^2}{\partial x^2}\b{U}_P(\b{x},\b{x}^*).
\end{multline}


\subsection{Numerical method: the method of {\it stresslet images}}\label{sec:Computation}

The fundamental singularities of Stokes flow may be used to derive a representation formula for the fluid flow in terms of singular boundary integrals(see \cite{pm87} and \cite{Pozrikidis92}). The presence of a nearby wall has been incorporated into various forms of the boundary integral formulation, for instance using regularized Stokeslets with their images \citep{adebc08}. A well-posed double layer form of the boundary integral formulation may be adapted for use near an infinite wall using image singularities of the stresslet, as suggested by \cite{sl12}. In this double-layer formulation with stresslet images, the fluid velocity is given by (see \cite{Pozrikidis92}): 
\begin{gather}
\b{u}(\b{x})=-\int_{S} \b{q}(\b{y})\cdot (\t{T}(\b{x},\b{y})+\t{T}^*(\b{x},\b{y}^*))\cdot \b{\hat{n}}(\b{y})\,dS+\frac{1}{8\pi}\left(\t{G}(\b{x},\b{x}_0)+\t{G}^*(\b{x},\b{x}^*)\right)\cdot \b{F},
\end{gather}
where $\b{\hat{n}}$ is the unit normal vector pointing into the fluid, $\b{y}$ is an integration variable over the body surface, $\b{q}(\b{y})$ is an unknown density,
\begin{gather}
\t{T}(\b{x,y})=-6\frac{(\b{x-y})(\b{x-y})(\b{x-y})}{|\b{x-y}|^5}
\end{gather}
is the stresslet singularity, a third-order tensor, and $\t{T}^*(\b{x},\b{y}^*)$ is the \tcb{associated image system, which is singular at the image point $\b{y}^*$ inside the wall and is given by the formula
\begin{gather}
\label{eq:stressletImage}
\t{T}_{ijk}^*(\b{x,y}^*)=  \frac{6\hat{X}_i X_jX_k}{|\bm X|^5}  +12x_3\frac{\beta_{ik}y_3 X_j +\beta_{ij}y_3 X_k-\delta_{jk}x_3\beta_{i\ell}X_\ell}{|\bm X|^5} -60x_3y_3 \beta_{i\ell} \frac{X_jX_kX_\ell}{|\bm X|^7},
\end{gather}
where $\beta_{ij} = \delta_{ij}-2\delta_{3i}\delta_{3j}$ is the reflection operator, $\bm y^* = \bm\beta \b{y}$ (and $\bm y = \bm\beta\b{y}^*$), $\b{X} = \b\beta(\b{x}-\bm{y}^*)$, and $\hat{\b{X}}=\b{x}-\b{y}$. The expression \eqref{eq:stressletImage} is the result of applying the Lorentz reflection (see \cite{kuiken96} or \cite{kk91}) to the original stresslet, with some manipulation.} The dimensionless force due to gravity acts at an angle $\beta$ relative to the wall, $\b{F}=\cos\beta\b{\hat{x}}-\sin\beta\b{\hat{z}}$ (the wall is parallel to gravity when $\beta=0$), and $\b{\t{G}}^*(\b{x},\b{x}^*)\cdot \b{F}=\cos\beta\, \b{G}_x^*(\b{x},\b{x}^*)-\sin\beta\, \b{G}_z^*(\b{x},\b{x}^*)$.

In the limit as the point $\b{x}$ tends towards a point on the boundary, $\b{x}\in S$, the no-slip boundary condition on the body surface provides an integral equation to be solved for $\b{q}$,
\begin{multline}
\b{U}+\b{\Omega}\times(\b{x}-\b{x}_0)=-\int_{S} \left(\b{q}(\b{y})-\b{q}(\b{x})\right)\cdot (\t{T}(\b{x},\b{y})+\t{T}^*(\b{x},\b{y}^*))\cdot \b{\hat{n}}(\b{y})\,dS\\
+\frac{1}{8\pi}\left(\t{G}(\b{x},\b{x}_0)+\t{G}^*(\b{x},\b{x}^*)\right)\cdot \b{F}.\label{eq: doublelayer}
\end{multline}
The integrand is finite with a jump at the singular point, $\b{x}=\b{y}$. Further investigation of the integral operator leads to relations between the velocities and the density $\b{q}$, closing the system:
\begin{gather}
\b{U}=-\frac{4\pi}{S_A}\int_{S}\b{q}(\b{x})\,dS,\ \ \b{\Omega}= -\sum_{m=1}^3\frac{4\pi}{A_m}\b{e}_m\left(\b{e}_m\cdot \int_{S}(\b{x}-\b{x}_0)\times\b{q}(\b{x})\,dS\right),
\end{gather}
where $S_A$ is the surface area of the particle, $\b{e}_m$ is the $m^{th}$ Cartesian unit vector, and $A_m=\int_{S} |(\b{e}_m\times (\b{x}-\b{x}_0))|^2\,dS$ (see \cite{Pozrikidis92}).

\tcb{To solve the integral equations above we use a collocation scheme, enforcing the equations at the nodes of the quadrature rule used to approximate the surface integrals. The body surface is parameterized using a spherical coordinate system. Integration is performed with respect to the zenith angle using Gaussian quadrature with $N_\phi$ points and with respect to the azimuthal direction using the trapezoidal rule with $N_\theta$ points, with $N_\theta$ varying with the dimensionless ring circumference $R\in(0,1)$ so as to achieve a roughly uniform distribution over the surface; namely, $N_\theta$ is taken to be the greatest integer not exceeding $N_\phi R$ in the prolate case or $2.5 N_\phi R^2$ in the oblate case. The integrand in Eq.~\eqref{eq: doublelayer} is set to zero at the jump discontinuity, a convenient method pioneered by \cite{pm87}, resulting in a quadrature scheme which is second-order accurate in the grid-spacing. Where time-stepping is required we employ a second-order Runge-Kutta method. The grid-spacing and time-step size are chosen based on the stiffness of the problem under investigation, and changes to the numerical results presented in the paper are negligible when compared to simulations with much finer resolution. A convergence study, comparisons to previously published numerical results, and other methods of validation are included as Appendix \ref{sec:convergence_tests}.}


A major benefit of the double-layer boundary integral method with stresslet images is that the equation for the density $\b{q}$ is a Fredholm integral equation of the second kind and is therefore well-posed, unlike approaches built upon a single-layer or combined formulation \citep{Pozrikidis92,sh11}. Moreover, the flow can be computed using adaptive quadrature for near-wall interactions without suffering from poor conditioning problems, and other subtle issues like the careful selection of a regularization parameter may be avoided. \tcb{The method of stresslet images used in this paper is more accurate than the popular method of regularized Stokeslets with images by \cite{adebc08} in the tests performed in Appendix \ref{sec:convergence_tests}, and does not rely on an extra regularization parameter.} A more complete discussion on the numerical method will be presented elsewhere.

\section{A zoology of particle trajectories}
\label{sec:NumTraj}
We begin by conducting a numerical survey of the trajectories exhibited by sedimenting prolate and oblate bodies near a vertical or tilted wall. \tcb{Figures \ref{Trajboard} and \ref{fig:AwesomeFigure}} show a selection of representative body dynamics, which depend on the body shape, initial data, and wall inclination angle. We presently discuss each trajectory type in turn. 

\subsection{Glancing, reversing, and tumbling near a vertical wall}

\begin{figure}
\centering
\includegraphics[width=.7\textwidth]{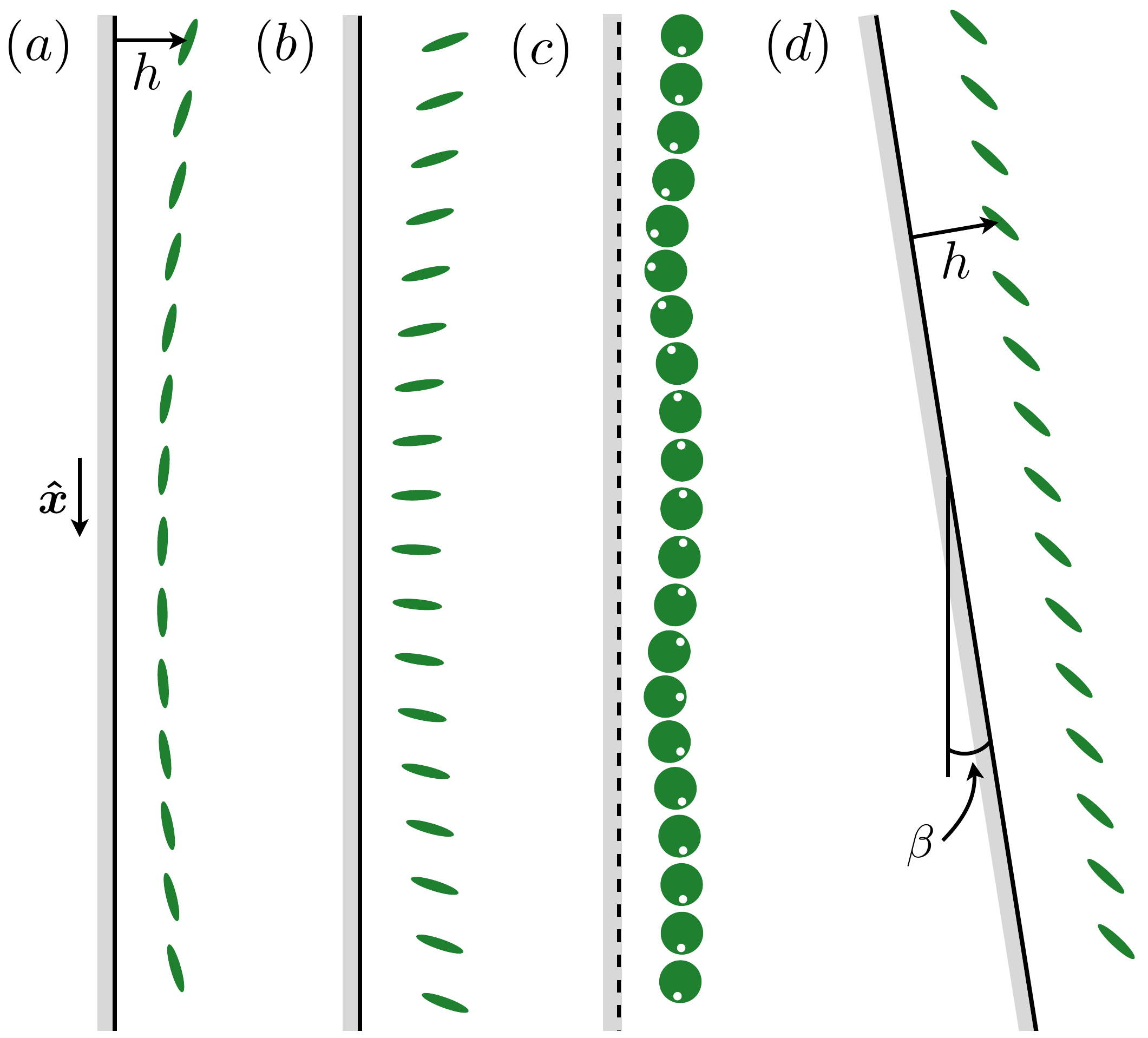}
\caption{Trajectories of prolate spheroidal bodies sedimenting near walls, as determined by full numerical simulation: (a) symmetric glancing; (b) symmetric reversing; (c) periodic tumbling (the distance traveled along the wall has been scaled by a factor of $100$); (d) stable sliding near an inclined wall.  The white markers on the tumbling body illustrate the rotation. Initial data and body shapes in each case are given in Appendix~\ref{appendix:initialconditions}.} 
\label{Trajboard}
\end{figure}
Our investigation begins in the simplest setting, where the wall is parallel to gravity, $\beta=0$, and the geometry has symmetry through the $xz$-plane, $\phi=0$. In this case, all trajectories can be described by tracking the distance $h$ from the particle center to the wall together with the angle $\theta$ measuring rotation in the $xz$-plane. Figure~\ref{Trajboard}a shows the {\it glancing} dynamics of a slender prolate spheroid of eccentricity $e=0.98$, placed initially at a distance $h=3$ from the wall and at an orientation angle $\theta=-20^\circ$. Due to the drag anisotropy of slender bodies in viscous flows, the body initially drifts towards the wall. Hydrodynamic interactions with the surface then cause the particle to rotate until $\theta=0$, at which point, in accordance with the time-reversal symmetry of the Stokes equations, the body continues to rotate and migrates away from the surface along a trajectory symmetric with its initial approach. As the particle escapes from the wall, its rotation rate diminishes so that $\theta$ tends toward a constant value. The same body follows a markedly different trajectory if the initial orientation angle is larger, as shown in Fig.~\ref{Trajboard}b, a {\it reversing} trajectory. In this example the initial orientation is $\theta\approx-70^\circ$, the body rotates in the opposite direction, and the leading edge becomes the trailing edge after the closest approach to the wall.

These glancing and reversing dynamics were explored numerically, analytically, and experimentally by \cite{rhlt77} for very slender particles. In the trajectories described in that work, a particle released far from the wall with $\theta\in(-\pi/2,0)$ always approaches the wall, rotates, and then escapes from the wall, just as shown in Fig.~\ref{Trajboard}a-b. \cite{rhlt77} distinguished glancing from reversing trajectories according to the orientation of the particle at closest approach to the wall; in the former the particle is oriented parallel to the wall at closest approach while in the latter the particle is oriented normal to the wall at closest approach. \tcb{The distinction between glancing and reversing can also be described in terms of a vector $\b{d}$ aligned with the axis of body symmetry: $\b{\hat{x}}\cdot \b{d}$ does not change sign in prolate glancing and oblate reversing orbits, while it does change sign in oblate glancing and prolate reversing orbits.}

A third trajectory type prevails for a nearly spherical body released near the wall, as shown in Fig.~\ref{Trajboard}c, where we take $e=0.15$. Releasing the body at $h=3$ with $\theta=0$, we observe a new type of dynamics, a \tcb{slow} periodic {\it tumbling} motion \tcb{(the distance traveled along the wall is scaled by a factor of $100$ in Fig.~\ref{Trajboard}c)}. \tcb{This behavior can be understood as a perturbation of the well-known trajectory of a sphere near a vertical wall, i.e. translation in the direction of gravity together with a rolling-type rotation due to the torque induced by the presence of the wall. The slight eccentricity of the body causes a drag anisotropy which in turn} leads to a migration velocity of the particle either toward or away from the surface, depending on the orientation angle. This slight migration, in concert with the rolling-type rotation, leads to a periodic tumbling trajectory. The dynamics are similar to the periodic tumbling of two identical non-spherical bodies placed side by side, as studied by \cite{Kim85,Kim86} and \cite{jspst06}, with an important difference: here the rotation of the body is in the opposite direction, $\dot{\theta}<0$ (relating to the opposing orientation of the Stokeslet in the image system in Eq.~\eqref{eq: Gstar}). 





\subsection{Sliding along an inclined wall}
Another type of particle trajectory arises when the bounding wall is not parallel to gravity. Figure~\ref{Trajboard}d shows the dynamics of a prolate spheroid with $e=0.98$ near a wall which is tilted at an angle $\beta=9.17^\circ$ (the initial data for the cases shown is included as \tcb{Appendix~\ref{appendix:initialconditions}}). Here a behavior appears that does not exist for sedimentation near a vertical wall, which we term {\it sliding}.  The body settles into a steady motion with a fixed orientation angle and distance from the wall. In this quasi-steady equilibrium the horizontal velocity induced by the particle orientation exactly balances the approach of the wall as the particle falls, and the rotation of the body due to the interaction with the wall has vanished; this type of trajectory was observed numerically by \cite{hg94}.  This quasi-steady state owes its existence to the breaking of gravity-wall symmetry in connection with the choice $\beta>0$, which weakens the consequences of time-reversibility on the dynamics.

Assuming $\phi=0$, the dynamics remain two-dimensional and it is natural to ask whether the glancing, reversing, and periodic tumbling trajectories found near a vertical wall still may be found near an inclined wall, and if so how they share the phase space with the sliding trajectory.  Although  not shown in Fig.~\ref{Trajboard}, the combination of small wall inclination angle $\beta$ and large eccentricity $e$ allows both glancing- and reversing-like trajectories to occur in the full numerical simulations. However, these trajectories are less symmetric in that the limiting orientation angle after the wall encounter is no longer the opposite of the value before the wall encounter for the same distance to the wall; instead, the wall interaction tends to focus the orientations of escaping particles into a narrow band of escape angles. As $\beta$ increases or $e$ decreases, \tcb{it becomes increasingly difficult for the particle to escape from the wall and the concentration of escape angles increases} until it \tcb{yields} an attracting fixed point, \tcb{namely} the sliding trajectory discussed above. For still larger $\beta$ the equilibrium particle-wall gap size becomes extremely small, resulting in excessive computational costs \tcb{and possible wall impact}, and we do not study this regime.  On the other hand, a careful tuning of $\beta$ against particle eccentricity can produce geometries where the fixed point is arbitrarily far from the wall, yet finite.  \tcb{In \S \ref{sec:SymbTraj} we derive analytical results quantifying this phenomenon, illustrated in Figure \ref{fig:H0vsBetaE}.}

The periodic tumbling orbits mentioned earlier no longer exist \tcb{with $\beta>0$}. Instead, a nearly spherical body is found to rotate in nearly periodic orbits, but with a slow drift toward the wall (for $\beta>0$) until eventually the orbit approaches the wall very closely.  These initially near-periodic trajectories may be of more mathematical interest than practical application, since the region in parameter space where they arise is so limited. 


\subsection{\tcb{Three-dimensional} glancing, reversing, \tcb{and wobbling}}\label{sec:3Dgrw}

\begin{figure}
\centering
\includegraphics[width=\textwidth]{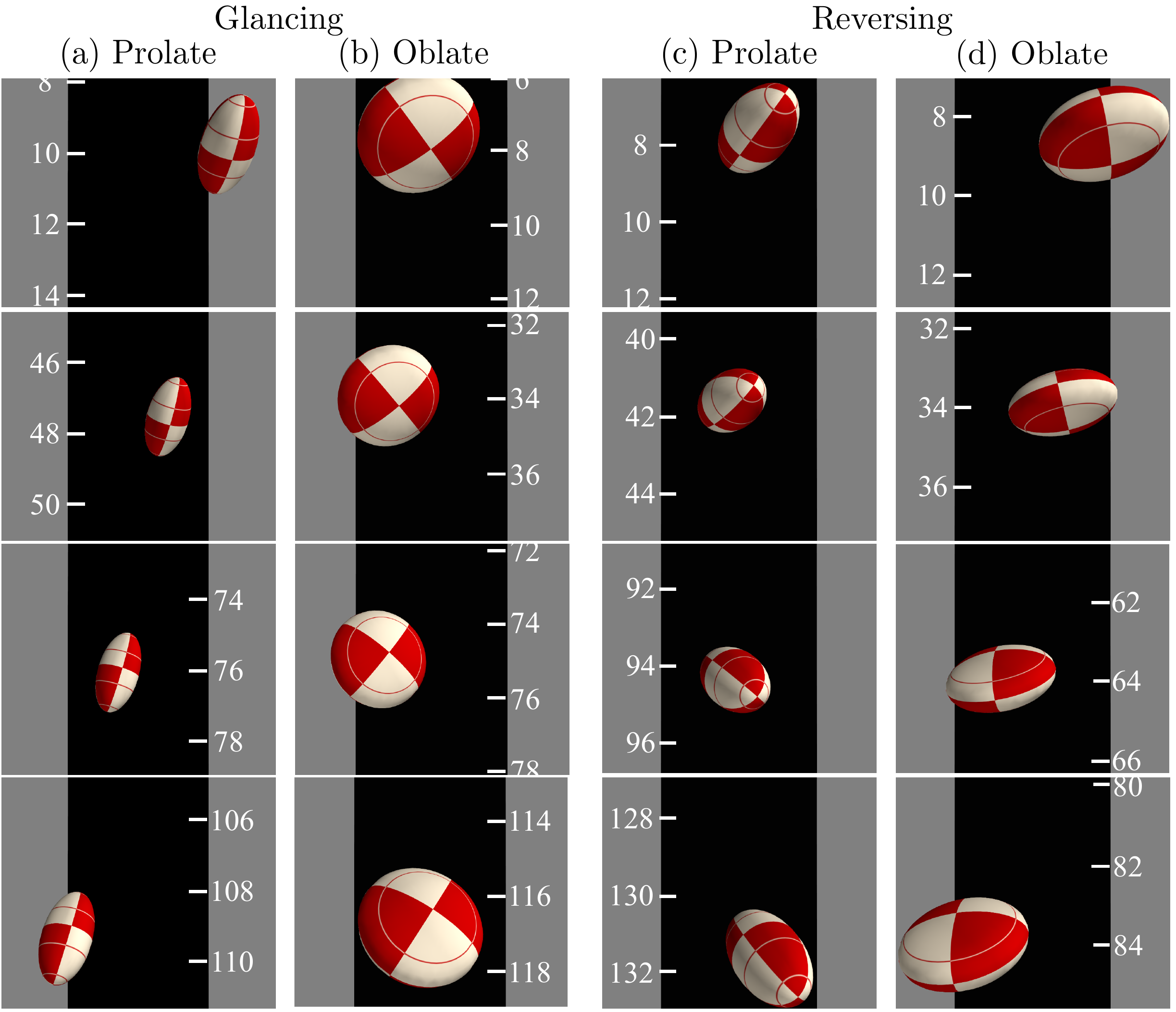}
\caption{Three-dimensional glancing and reversing of prolate and oblate bodies near a vertical wall.  
The black rectangle in the background of each frame represents a strip of the wall, $\{(x,y,0):-2\le y\le 2\}$.  
Gravity is parallel to the wall, i.e. vertical on the page; the horizontal axis is the $y$-direction.
The lateral movements are plotted to scale, while the movements in the $x$-direction have been greatly reduced for visualization purposes.  
Animations of these four trajectories are included as supplementary material.  The initial data used to generate these trajectories is given in Appendix~\ref{appendix:initialconditions}. The movies, along with movies of periodic tumbling and wobbling of nearly-spherical prolate and oblate bodies, are included as supplementary material.}
\label{fig:AwesomeFigure}
\end{figure}

In general a particle near a surface will undergo lateral translations and out-of plane rotations leading to a fully three-dimensional trajectory. Consider a particle falling near a vertical wall, $\beta=0$, but with no lateral symmetry, $\phi\neq 0$. \tcb{Four such trajectories are shown in Fig.~\ref{fig:AwesomeFigure}, where prolate and oblate spheroids with aspect ratio $a/c=2$ have been released with nonzero values of both $\theta$ and $\phi$.  A movie depicting these trajectories is included as supplementary material.} Depending on the initial data, we again observe glancing- and reversing-like trajectories wherein the particle approaches and then escapes from the wall; \tcb{in three-dimensions, glancing and reversing trajectories can be classified just as in the two-dimensional case, in terms of a vector $\b{d}$ aligned with the axis of body symmetry. Once again, $\b{\hat{x}}\cdot \b{d}$ does not change sign in prolate glancing and oblate reversing orbits, while it does change sign in oblate glancing and prolate reversing orbits.} Unlike in the two-dimensional case, the particle can drift laterally, subject to the constraint of symmetry about the moment of closest approach of the centroid to the wall.  \tcb{Importantly, the lateral behavior depends on whether the particle is prolate or oblate. A glancing prolate body and a reversing oblate body continue drifting laterally without a change in the direction of drift (Figs.~\ref{fig:AwesomeFigure}a\&d), while a glancing oblate body and a reversing prolate body return in the direction from which they came at the point of closest approach (a sign change in $\dot{y}$, Figs.~\ref{fig:AwesomeFigure}b\&c).} 

The periodic tumbling trajectory \tcb{also has fully three-dimensional analogues. Included in the supplementary material is a movie showing the nearly-spherical prolate tumbling and oblate tumbling} \tcb{trajectories which have a periodic lateral wobble with zero net lateral drift. These tumbling orbits, rotated away from the two-dimensional dynamics previously described, are now found to undergo periodic lateral motions in the $y$ direction. As in the two-dimensional case the trajectory can be understood as a combination of spherical rolling and reversing. The difference in prolate and oblate lateral drift in Fig.~\ref{fig:AwesomeFigure}c\&d also emerges in the three-dimensional tumbling orbits, so that the body changes lateral direction at the point of closest approach in the prolate case and at the point farthest from the wall in the oblate case.}

In the more general \tcb{setting} with $\phi\neq0$ and $\beta>0$ we have observed in numerical simulations that for small $\beta$ (small wall tilt angle) the wall interactions induce a concentration of the three-dimensional dynamics (escape angles tend toward a narrower band). For larger values of $\beta$ we see the emergence of an attracting fixed point. The wall inclination damps $\phi$ towards 0, and the fixed point is the same as in the case of lateral symmetry as illustrated in Figure \ref{Trajboard}d. \tcb{We will return to all of these trajectory types once we have developed analytical expressions with which to study them.}

\section{The method of images for wall-bounded Stokes flow}
\label{sec:moi}
The numerical investigations described above are somewhat computationally intensive; for each time step in a trajectory a large linear system representing the discrete version of the surface integral equation must be inverted. At the same time, the dynamics can be fully described by tracking two or three scalar parameters, and the derivation of ordinary differential equations describing their dynamics would be of considerable value. To obtain an explicit system of differential equations which can be rapidly integrated or further studied analytically, we will apply the method of images and the method of reflections to an arbitrarily oriented prolate or oblate spheroid near a vertical or inclined wall.

The method of reflections takes an especially convenient form when the flows are constructed from systems of fundamental singularity solutions of the Stokes equations. The flow due to the motion of an spheroidal body in an infinite fluid may be represented by a collection of singularities placed at points interior to the body surface; image systems are then placed at the reflections of these points inside the wall to enforce the no-slip boundary condition on the wall. A generalization of Fax\'en's Law then gives the effect of this auxiliary velocity field on the body as a first-order correction of the trajectory due to the wall. The process may be continued to develop higher-order approximations of the effect of the wall on the body trajectory (see \cite{kk91}). \cite{Wakiya59} carried out a similar procedure using Lamb's solutions in ellipsoidal coordinates, producing expressions for the force and torque on a body moving with lateral symmetry near a wall.

The flow field associated with an spheroidal body in an unbounded fluid may be represented by an integrated distribution of image singularities on the centerline (prolate case) or a circular disk (oblate case). However, far from the particle, $r=|\b{x}-\b{x}_0| \gg 1$, this velocity field may be written as a multipole expansion of singularities placed at the body centroid. As shown in \cite{kk91}, the dimensionless fluid velocity far from the body is given by
\begin{gather}
\label{eq:ME}
\b{u}^{(0)}(\b{x}) = \frac{3}{4}\left(\frac{\sinh(D)}{D}\right)\t{G}(\b{x},\b{x}_0)\cdot \b{F},
\end{gather}
where $\b{F}=\cos\beta\b{\hat{x}}-\sin\beta\b{\hat{z}}$ is the external gravitational force on the body, $\t{G}$ is the Stokeslet singularity given in Eq.~\eqref{eq: Stokeslet}, and
\begin{gather}\label{Eq: sinhD}
\frac{\sinh(D)}{D}=\sum_{n\ge 0} \frac{1}{(2n+1)!}D^{2n}=1+\frac{1}{6}D^2+\cdots,
\end{gather}
with $D^2 = \partial_{XX}+(b/a)^2\partial_{YY}+(c/a)^2 \partial_{ZZ}$. Truncating the series after the two terms shown above results in errors in the flow (from Eq.~\eqref{eq:ME}) that scale as $r^{-5}$ as $r\rightarrow \infty$. It will prove useful to transform the differential operator $D^2$, which is diagonalized in the coordinate system of the spheroidal body axes, into the usual coordinate system oriented \tcb{with the wall at $\{z=0\}$}. Given the definitions of $\theta$ and $\phi$ from Eq.~\eqref{eq:Geometry} the second derivatives can be written as linear combinations of derivatives along the standard axes,
\begin{align}\label{Eq: dXXs}
\begin{split}\partial_{XX}=& 
\cos^2\theta\cos^2\phi\, \partial_{xx}+\cos^2\theta\sin^2\phi\, \partial_{yy}
+\sin^2\theta\, \partial_{zz}\\
&+\sin(2\theta)\left(\cos\phi\, \partial_{xz}+\sin\phi\, \partial_{yz}\right)+\cos^2\theta\sin(2\phi)\,\partial_{xy},\end{split}\\
\partial_{YY}=& \sin^2\phi\,\partial_{xx}+\cos^2\phi\,\partial_{yy}
-\sin(2\phi)\,\partial_{xy},\\
\begin{split}\partial_{ZZ}=& 
\sin^2\theta\cos^2\phi\,\partial_{xx}+\sin^2\theta\sin^2\phi\,\partial_{yy}+\cos^2\theta\,\partial_{zz}\\
&-\sin(2\theta)\left(\cos(\phi)\,\partial_{xz}+\sin\phi\,\partial_{yz}\right)+\sin^2(\theta)\sin(2\phi)\,\partial_{xy}.\end{split}
\end{align}
Eq.~\eqref{eq:ME} may now be expressed in terms of the $x$- and $z$- directed Stokeslets and selected second derivatives.  To obtain an image system, we employ Blake's image system for the Stokeslet, and expressions such as Eq.~\eqref{eq:Sx_xzimage} for the required second derivatives. This leads to an analytical expression for the reflection flow, $\b{u}^{(1)}(\b{x})$. Since we have truncated the series in Eq.~\eqref{Eq: sinhD}, neglecting terms of size $D^4(r^{-1})$, the error in this reflected flow scales as $(r^*)^{-5}$ for $r^*\rightarrow \infty$, where $r^*=|\b{x}-\b{x}^*|$ and $\b{x}^*=(0,0,-h)$.

Finally, the effect of the reflected flow on the original spheroidal particle is given by the mobility relations between the particle motion $(\b{U},\b{\Omega})$ and the external force and torque $(\b{F},\b{T})$ through a generalized Fax\'en law (see \cite{kk91}),
\begin{align}
-\b{F} &= (X^A\b{d}\b{d}^T+Y^A(I-\b{d}\b{d}^T))\cdot \left(\frac{\sinh(D)}{D}\b{u}^{(1)}(\b{x}_0)-\b{U}\right),  \label{eq:FL1}\\
\nonumber-\b{T} &= \frac{2}{3}(X^C\b{d}\b{d}^T+Y^C(I-\b{d}\b{d}^T))\cdot\left.\left(\frac{3}{D}\frac{\partial}{\partial D}\left(
\frac{\sinh(D)}{D}\right)\nabla\times\b{u}^{(1)}\right|_{\b{x}_0}-2\bm\Omega\right)\\
&\quad -\frac{4}{3}Y^H\left\{
\frac{3}{D}\frac{\partial}{\partial D}\left(\frac{\sinh(D)}{D}\right)\cdot \b{E}^{(1)}(\b{x}_0)\cdot \b{d}\right\}\times\b{d},
\label{eq:FL2}
\end{align}
where $\b{E}^{(1)}=(\nabla\b{u}^{(1)}+[\nabla\b{u}^{(1)}]^T)/2$ is the symmetric rate-of-strain tensor and $\b{d}$ is a unit vector oriented along the particle's axis of symmetry, with
\begin{gather}\label{eq:dsymmetry}\b{d} =\begin{cases}
(\cos\theta\cos\phi,\cos\theta\sin\phi,\sin\theta) &\textnormal{for prolate bodies,} 
\\(-\sin\theta\cos\phi,-\sin\theta\sin\phi,\cos\theta) &\textnormal{for oblate bodies}. 
\end{cases}\end{gather}
The constants $X^A, Y^A, X^C,Y^C$, and $Y^H$ depend only on the eccentricity $e$ and whether the particle is prolate or oblate, and are included in Table \ref{table:parameters}. The dimensionless torque $\b{T}$ is the result of scaling upon $a |\b{F}^*|$. The prolate and oblate problems are solved together in a single calculation through the introduction of a parameter which is equal to $1$ for prolate bodies and $-1$ for oblate bodies. Each component of $\b{dd}^T$ can then be rewritten using appropriate trigonometric identities; for example, $d_1d_1$ reduces to $\cos^2\phi(1\pm\cos(2\theta))/2$.  

Setting $\b{T}=0$ in Eq.~\eqref{eq:FL2} to solve the sedimentation problem of interest and truncating the differential operators in Eqs.~\eqref{eq:FL1}-\eqref{eq:FL2} at second order using Eq.~\eqref{Eq: sinhD} and 
\begin{gather}
\frac{3}{D}\frac{\partial}{\partial D}\left(\frac{\sinh(D)}{D}\right) = 1+\frac{D^2}{10}+\cdots ,
\end{gather}
we obtain linear equations for $\bm U$ and $\bm \Omega$. \tcb{Solving those equations and extracting $\dot{\theta}$ and $\dot{\phi}$ from $\bm{\Omega}$ results in a set of ordinary differential equations governing the dynamics of the body.  The full system, for an inclined wall, is given in Appendix A together with the formula for $\bm{\Omega}$.} When the wall is vertical, the system reduces to an especially tidy form:
\begin{align}
\dot{x}&=\frac{{\left({2-\cos^2\phi\left(1\pm \cos\left(2 \, \theta\right) \right)}\right)} {\left(X^A-Y^A\right)} +2 Y^A}{2 \, X^AY^A} 
-\frac{9}{16 \, h} \label{eq:xprimeV}\\ \nonumber
&\quad+ \frac{2 \, e^{2} {\left( \cos\left(2 \,\theta\right)\pm1\right)} \cos^2\phi + 18 \, e^{2}\cos^2\theta - e^{2} {\left(17\pm7\right)} +16}{128 \, h^{3}},\\
\label{eq:yprimeV}\dot{y}&= 
-\frac{\sin(2\phi){\left(1\pm \cos\left(2 \, \theta\right)\right)}{\left(X^A - Y^A\right)}}{4 X^A Y^A} 
+ \frac{e^{2}\sin(2\phi) {\left(\cos\left(2 \, \theta\right)\pm1\right)}}{128 \, h^{3}},\\
\label{eq:hprimeV}
\dot{h}&=  \frac{ \pm  \cos\phi\sin\left(2   \theta\right)  {\left(Y^A - X^A\right)}}{2   {X^A}Y^A } 
- \frac{e^{2} \cos\phi \sin\left(2   \theta\right)}{32   h^{3}}, \\
\label{eq:thetaprimeV}\dot{\theta}&= \frac{9\tcb{e^2} \cos\phi\cos\left(2   \theta\right) }{32 {\left(2 - e^{2} \right)}h^{2} }
\\&\quad-\frac{ 3\cos\phi}{64   {\left(2-e^{2}\right)} h^{4} }\cdot\Big[ 4 - 10 e^2 
+{\left(7\pm1\right)}e^{4}  + e^{2} \cos^2\theta {\left(9   e^{2} \cos^2\theta- {\left( 15\pm 2\right)}e^{2}  + 12\right)}\Big],\nonumber\\
\label{eq:phiprimeV} \dot{\phi}&=
\begin{cases}
\displaystyle\frac{3\sin\phi\tan\theta}{64 {\left(2-e^{2} \right)}}
\left(\frac{-6  e^{2} }{ h^{2}} + \frac{3  e^{4}
\cos^2\theta - 8  e^{4} + 10  e^{2} - 4}{ h^{4} }\right) & \textnormal{(prolate)},\\
\displaystyle\frac{3\sin\phi\cot\theta}{64 {\left(2-e^{2} \right)}}
\left(\frac{-6  e^{2} }{h^{2}} - \frac{3  e^{4}\sin^2\theta - 2  e^{4} - 2  e^{2} - 4}{ h^{4} }\right)&\textnormal{(oblate)},
\end{cases}
\end{align}
where the $\pm$ signs should be replaced with $+$ in the prolate case and $-$ in the oblate case, and the constants $X^A$ and $Y^A$ have different definitions in the two cases as indicated in Table~\ref{table:parameters}. Importantly, the derivatives of the particle-wall distance $h$ and of the angles $\theta$ and $\phi$ are independent of the positions $x$ and $y$, so that the system is fundamentally three-dimensional; the positions $\{x(t),y(t)\}$ may be determined directly once $\{h(t), \theta(t), \phi(t)\}$ have been found. The errors in the expressions above and in the general setting (in Appendix A) are $\mathcal{O}(h^{-4})$ in the translational velocity and $\mathcal{O}(h^{-5})$ in the rotational velocity for $h \gg 1$. 

The full expression of the rotational velocity $\bm{\Omega}$ could be used to deduce the rotation of the body about its axis of symmetry, a third angle that in addition to $\theta$ and $\phi$ prescribes the precise history of the body as it evolves in time. However, the translational and rotational velocities computed at any moment are invariant to rotations about the axis of symmetry, so that this third angle may be determined after solving for $\{h(t), \theta(t), \phi(t)\}$ just as may be done for the drift positions $x(t)$ and $y(t)$. 

\begin{table}
\centering \small
\begin{tabular}{c c c}
\toprule 
  & Prolate & Oblate\\
\midrule
$X^A$& $8e^3/\left[-6e+3(1+e^2)K\right] $ & 
$4e^3/\left[ (6e^2-3)K+3e\sqrt{1-e^2}\right] $ \\
$Y^A$& ${16}e^3/\left[6e+(9e^2-3)K\right] $ & 
$8e^3/\left[ (6e^2+3)K-3e\sqrt{1-e^2}\right] $ \\
$X^C$& $4e^3(1-e^2)/\left[6e-(3-3e^2)K\right] $ & 
$2e^3/\left[3K-3e\sqrt{1-e^2}\right] $ \\
$Y^C$& $4e^3(2-e^2)/\left[-6e+(3+3e^2)K\right] $ & 
$2e^3(2-e^2)/\left[ (6e^2-3)K+3e\sqrt{1-e^2}\right] $ \\
$Y^H$& $4e^5/\left[-6e+(3+3e^2)K\right] $ & 
$-2e^5/\left[(6e^2-3)K+3e\sqrt{1-e^2}\right] $ \\
$K$&$\log\left([1+e]/[1-e]\right)$&$\arccot\left(\sqrt{1-e^2}/e\right)$\\
$\pm$&$+$&$-$\\
\bottomrule
\end{tabular}
\caption{Parameter definitions for the prolate and oblate cases in the Fax\'en laws \eqref{eq:FL1}-\eqref{eq:FL2} and the general system \eqref{eq:xprime}-\eqref{eq:phiprime}, from \cite{kk91}.  Of these, only $X^A$ and $Y^A$ appear directly in \eqref{eq:xprime}-\eqref{eq:phiprime} because we have exploited the relation $Y^H/Y^C = \pm e^2/(2-e^2)$. The quantity $\pm(X^A-Y^A)$ is negative for both body types. }
\label{table:parameters} 
\end{table}




\section{Analysis of particle trajectories}
\label{sec:SymbTraj}
The ordinary differential equations describing the body dynamics can be integrated numerically, and in some cases analytically, to produce approximate trajectories for the sedimentation problem in the general setting. In this section we will derive analytical formulae for the particle trajectory in various special cases, beginning with the assumption of two-dimensional dynamics and then proceeding to the general case.  
\subsection{Analysis of glancing, reversing, and tumbling dynamics}\label{sbs:GRT}
Consider first the case of two-dimensional motion, $\phi=0$, near a vertical wall, $\beta=0$. The evolution of the particle position and orientation is governed by the reduced system (from Eqs.~\eqref{eq:hprimeV} and \eqref{eq:thetaprimeV}),
\begin{align}
\label{eq:thetaABCDEF} \dot{\theta}&=\frac{\cos(2\theta)}{h^2}\left[ A- \frac{B}{h^2}- C\frac{\cos(2\theta)}{h^2}\right] - \frac{D}{h^4},\\
\label{eq:hABCDEF} \dot{h}&=\sin(2\theta)\left[E - \displaystyle\frac{F}{h^3}\right],
\end{align}
where 
\begin{gather}
\label{eq:ABCDEF}
\begin{split}
A=\frac{9 e^2 }{32 (2-e^2)}, \ \  
B= \frac{3 e^{2} {\left(6-{\left( 3\pm 1\right)}e^{2}  \right)} }{64  {\left(2-e^{2} \right)}}, \ \ 
C=\frac{  27 e^4 }{256 (2-e^2) }, \\
D= \frac{48-48 e^2+21 e^4}{256 (2-e^2)  }, \ \  
E= \pm\frac{Y^A-X^A }{2X^AY^A},\ \ 
F=  \frac{e^{2}}{32}.
\end{split}
\end{gather}
The limiting case of a spherical body is dramatically simpler, with $A=B=C=E=F=0$, and $D=3/32$. For a general particle eccentricity $e$ the system has a fixed point at $\theta=0$ and a particle-wall distance that satisfies
\begin{gather}
\label{eq:unstablefixedpoint}
h^2= \frac{B+C+D}{A} =\frac{4+2e^2-(-1\pm1)e^4}{6e^2}.
\end{gather}

\begin{figure}
\centering
 \includegraphics[width=0.65\textwidth]{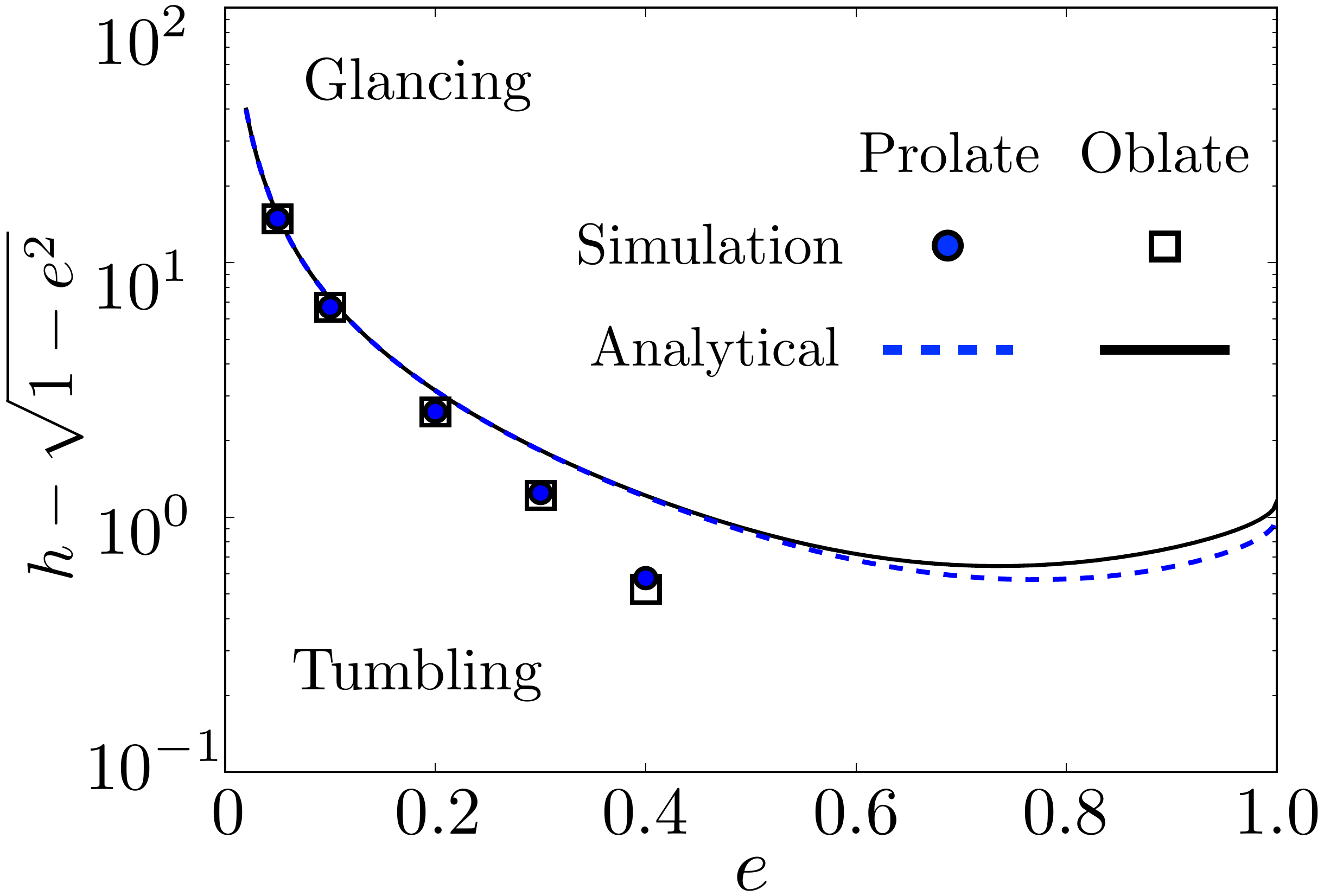}
\caption{The gap size for which a body with $\theta=0$ does not rotate as a function of particle eccentricity, which distinguishes the transition from glancing to periodic tumbling. The results of full numerical simulations are shown as symbols and those according to Eq.~\eqref{eq:unstablefixedpoint} as lines. The results for prolate and oblate bodies are remarkably similar, with the accuracy of the analytical estimates for both deteriorating for large particle eccentricity where the equilibrium distance is very close to the wall.} 
\label{fig:unstablefixed}
\end{figure}

\tcb{This unstable fixed point corresponds to a particle aligned with the wall and falling vertically without rotating, and can occur only at a specific particle-wall distance where the competing dynamics that give rise to glancing and reversing precisely balance each other. The corresponding particle-wall gap size $h-\sqrt{1-e^2}$ from \eqref{eq:unstablefixedpoint} is plotted against particle eccentricity in Fig.~\ref{fig:unstablefixed}, along with the values computed using the full numerical simulation, shown as symbols. The equilibrium distance is unbounded as $e\to 0$ (a sphere always rotates in the same direction at any finite distance from the wall). As particle eccentricity increases, the numerically determined gap size decreases and then vanishes, and the accuracy of the estimate from \eqref{eq:unstablefixedpoint} is poor for large particle eccentricity where the particle equilibrium distance is very close to the wall.}

A reduced but analytically tractable approximation of the system above is found in the limit of large particle distances from the wall and upon inspection of the coefficients. It is appealing to keep only the terms of size $O(h^{-2})$ in Eq.~\eqref{eq:thetaABCDEF} and of size $O(1)$ in  Eq.~\eqref{eq:hABCDEF}, but higher order terms become dominant when $\cos(2\theta)=0$. Moreover, for nearly spherical particles, $e\approx 0$, a more appropriate comparison of terms involves the ratio $e/h^2$; for instance, $B/h^4\sim e^2/h^4 \ll D/h^4$ for $e\ll 1$ and $h \gg 1$. Neglecting the terms with coefficients $B,C$, and $F$ results in the reduced system,
\begin{gather}
\dot{\theta}=\frac{A\cos(2\theta)}{h^2} - \frac{D}{h^4}, \ \ \ \ \dot{h}=E\sin(2\theta).
\end{gather}
This system is autonomous, and \tcb{the two derivatives can be divided to obtain a single first-order equation for $d\theta/dh$}. The transformations $\u=1/h$ 
and
$\eta = -\cos(2\theta)/2$ then yield a linear differential equation,
\begin{equation}
\frac{d\eta}{d\u} = \frac{2A}{E}\eta+\frac{D}{E}\u^2.
\end{equation}
Multiplying by $\exp\left({-2A\u/E}\right)$ and integrating leads to
\begin{gather}
\eta \exp\left({-\frac{2A}{E}\u}\right)=
 \frac{D}{E}\exp\left({-\frac{2A}{E}\u}\right)\left[
\frac{-E}{2A}\u^2
-\frac{E^2}{2A^2}\u
-\frac{E^3}{4A^3}   \right]  +c_0,
\end{gather}
where $c_0$ is a constant of integration. For each trajectory the relation
\begin{equation}
2c_0 = \exp\left({-\frac{2A}{E}\u}\right)\left(2\eta +\frac{D}{A}\left(\u^2+\frac{E}{A}\u+\frac{E^2}{2A^2}\right)\right)
\end{equation}
holds, and therefore each trajectory must follow a level set of the function 
\begin{equation}
\label{eq:Psi} \Psi(h,\theta) = \exp\left({-\frac{2A}{Eh}}\right)\left(-\cos(2\theta)+\frac{D}{A}\left(h^{-2}+\frac{E}{Ah}+\frac{E^2}{2A^2}\right)\right)
\end{equation}
in the $\theta h$-plane.  

\begin{figure}
\centering
 \includegraphics[width=\textwidth]{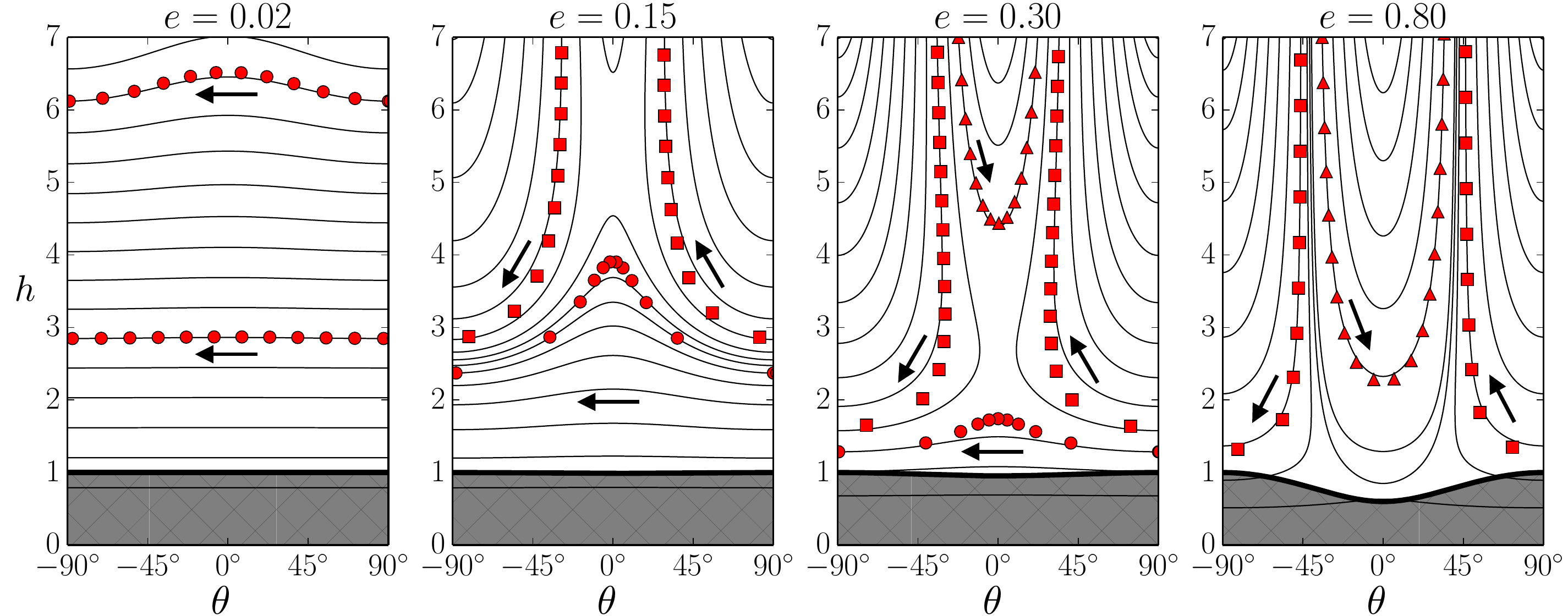}
\caption{Two-dimensional trajectories of prolate spheroids sedimenting near a vertical wall are depicted by plotting the particle-wall distance $h$ against the orientation angle $\theta$. \tcb{Unphysical regions with $h\leq \sqrt{\sin^2\theta+(1-e^2)\cos^2\theta}$, corresponding to body penetration through the wall, are shaded.} The results of the full numerical simulations are shown as red symbols. For small body eccentricity $e$ we observe periodic orbits near the wall (circles).  As $e$ increases, the periodic trajectories are replaced by reversing (squares) and glancing (triangles) trajectories. Arrows indicate the direction of time. The contours of the scalar function $\Psi$ from Eq.~\eqref{eq:Psi}, shown as black lines, give accurate predictions of the full numerical results.}
\label{fig:5traj}
\end{figure}

Figure~\ref{fig:5traj} shows the level sets of Eq.~\eqref{eq:Psi} as black lines, together with the results of the full numerical simulations (see \S\ref{sec:Computation}) as red symbols, for prolate spheroids of four different eccentricities: $e\in\{0.02,0.15,0.3,0.8\}$.  Periodic tumbling orbits are indicated by circles, reversing trajectories by squares, and glancing trajectories by triangles. Arrows indicate the direction of time. For $e=0.02$ the particle is nearly spherical and we see the periodic orbits described earlier. In these orbits the particle is farthest from the wall when $\theta=0$, i.e. when the major axis is parallel to the wall. \tcb{As noted in \S\ref{sec:NumTraj}, the period of the tumbling orbit is extremely long; using \eqref{eq:thetaABCDEF} for the limiting case of a sphere, $e\to 0$, we find the period of full rotation $T=64\pi h_0^4/3$, where $h_0$ is the constant distance from the wall. The sedimentation distance $X$ traveled during one period of a tumbling orbit, using \eqref{eq:xprimeV}, is}
\begin{gather}
\tcb{X = \frac{64\pi h_0^3}{3}\left(1-\frac{9}{16 h_0}+O\left(h_0^{-3}\right)\right)}.
\end{gather}

For eccentricities $e=0.15$ and $0.3$ in Fig.~\ref{fig:5traj}, the periodic orbits are restricted to a narrower region near the wall while glancing and reversing trajectories approach ever nearer to the wall before turning back.  Finally, with $e=0.8$, we see only glancing and reversing type trajectories, in concurrence with the slender-body work in \cite{rhlt77}; the left-most plot of Fig. \ref{fig:5traj} can be compared directly to Fig. 4 in that work, bearing in mind that the $\theta$ defined therein is the same as our $-\theta$.  

These contours concur with the numerical survey in classifying the trajectories in this symmetric version of the problem into glancing, reversing, and tumbling types.  The quantitative agreement with the numerical solutions is \tcb{generally very good, though imperfect in some cases where the body comes very close to the wall}. Our experience suggests that the results of the reduced equations should be used with some caution for $h<2$, \tcb{though the dynamics generally remain qualitatively sound for much smaller values of $h$}. Replacing the contours of $\Psi$ with trajectories determined by numerical integration of the fourth-order system \eqref{eq:thetaABCDEF}-\eqref{eq:hABCDEF} \tcb{results in only minor changes.}

\begin{figure}
\centering
\includegraphics[width=.7\textwidth]{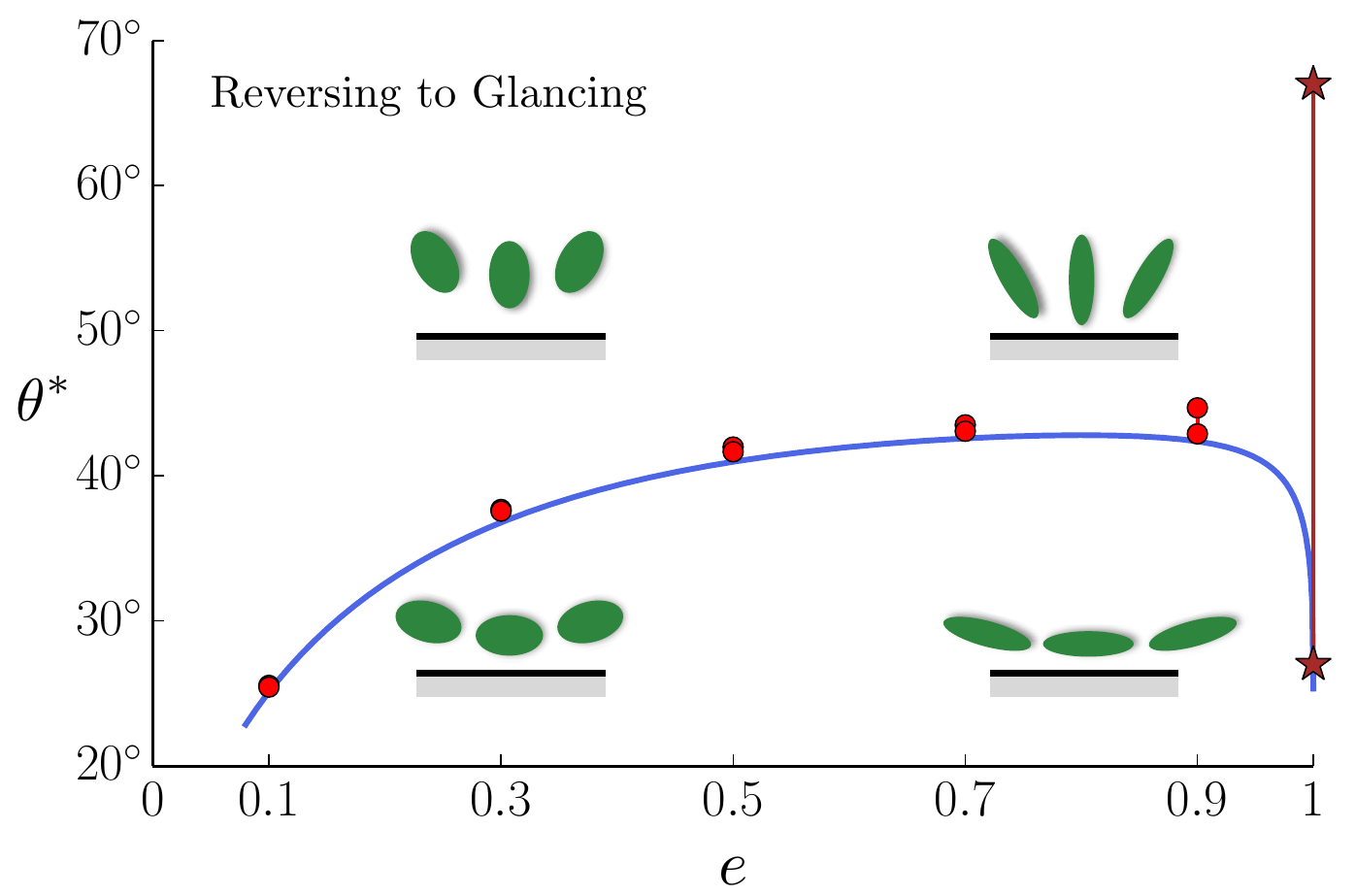}
\caption{The transition angle between glancing and reversing trajectories for prolate bodies as a function of eccentricity.  The solid curve is from the explicit formula \eqref{eq:ThetaStar}. Numerical results are shown as red circles; at low eccentricity the transition angle is well resolved but at greater eccentricities we report only a range of possible values. The two stars at right indicate the maximum glancing angle and the minimum reversing angle reported by \cite{rhlt77}, who in addition to a numerical study used aluminum wires of aspect ratio $a/c\approx60$ or $e=0.999861$.  Between these angles they reported wall impacts.  \tcb{For oblate bodies, the theory predicts a slightly larger value of $\theta^*$ than in the prolate case, but the difference is less than one degree for $e<0.866$.  }}
\label{fig:Glance-Reverse}
\end{figure}

From the analytical picture above we are in a position to predict solely from initial conditions whether or not a particle will escape, and if so to predict the final orientation it will assume far from the wall. The initial data $h=h_0$ and $\theta=\theta_0$ determine a level set of $\Psi$, and if the particle escapes this contour must have an asymptote with $h\rightarrow \infty$.  In this limit we obtain from Eq.~\eqref{eq:Psi} the relation
\begin{equation}
\label{eq:LimitThetaPsi} \cos(2\theta) = \frac{DE^2}{2A^3} - \Psi(h_0,\theta_0).
\end{equation}
If the right-hand side of \eqref{eq:LimitThetaPsi} has magnitude greater than one, there is no solution and a periodic orbit is predicted.  Otherwise, \eqref{eq:LimitThetaPsi} predicts the \tcb{limiting} orientation angle that the particle takes once it has escaped from the wall. Among escaping particles we can distinguish the glancing from the reversing trajectories by examining this asymptotic orientation angle more closely. That is, we consider the problem of determining the angle $\theta^*$ which divides glancing from reversing trajectories at a given eccentricity far from the wall. This can be done analytically by solving the equation $\Psi(\tcb{\sqrt{D/A}}, 0) = \lim_{h\rightarrow\infty} \Psi(h,\theta^*)$, since $(h=\tcb{\sqrt{D/A}},\theta=0)$ is the fixed point in the reduced model used to derive $\Psi$ 
and the glancing-reversing separatrix passes through this fixed point. \tcb{This gives an expression for $\theta^*$ in terms of the constants $A,D,E$ defined in \eqref{eq:ABCDEF}:
\begin{gather}\label{eq:ThetaStar}
\theta^*=\frac12\arccos\left(2\kappa^{-2} \left(1-\frac{\kappa +1}{\exp(\kappa)}\right)\right),\quad\tcb{\textnormal{where}\quad\kappa = \frac{2A^{3/2}}{E\sqrt{D}}.}
\end{gather}
Making the substitutions in \eqref{eq:ABCDEF} and in Table \ref{table:parameters}, we can write $\kappa$ explicitly in terms of the eccentricity $e$:
\begin{gather}
\kappa_{\textnormal{prolate}} = \frac{12 \sqrt{6} e^6}{\left(2-e^2\right) \sqrt{16-16 e^2+7 e^4} \left(\left(3-e^2\right) \log\left((1+e)(1-e)^{-1}\right)-6 e\right)},
\\
\kappa_{\textnormal{oblate}} = \frac{6 \sqrt{6} e^6}{\left(2-e^2\right) \sqrt{16-16 e^2+7 e^4} \left(\left(3-2e^2\right) \arccot\left(\sqrt{1-e^2}/e\right)-3e\sqrt{1-e^2}\right)}.
\end{gather}
}
To assess the accuracy of this analytical result we consider prolate spheroids and determine $\theta^*$ numerically for several values of $e$ by computing trajectories which start from a fixed large $h$ and various initial angles $\theta\in(-\pi/2,0)$ and continue until either $\theta$ reaches $0$ (glancing) or $-\pi/2$ (reversing). These numerical results are not obtained by integrating \eqref{eq:hprimeV}-\eqref{eq:thetaprimeV} but by solving the full Stokes equations using the numerical method described in \S\ref{sec:Computation}. The two values of the initial angle where the outcome changes from glancing to reversing determine an interval containing $\theta^*$. These results are shown in Fig.~\ref{fig:Glance-Reverse}. For $e\ll1$ the interval of uncertainty is quite small, and we report $\theta^*$ to an accuracy of $0.1^\circ$. For $e\approx1$ the trajectories of interest pass extremely close to the wall, resulting in excessive computational costs, so the intervals to which we are able to constrain $\theta^*$ are large enough to be visible in Fig.~\ref{fig:Glance-Reverse}. In an experimental setting, trajectories with initial angles within these ranges of uncertainty may result in wall collisions due to imperfections in the particle or wall geometry. The stars in Fig.~\ref{fig:Glance-Reverse}, corresponding to an eccentricity $e=0.99986$, are the values reported by \cite{rhlt77} as the result of numerical work and experiments with aluminum wires of aspect ratio $60$. The results compare well; for $e\in\{0.1,0.3,0.5,0.7,0.9\}$ the formula \eqref{eq:ThetaStar} gives a result within one degree of the numerical range, and for $e=0.999861$ the result is within one degree of the range reported by \cite{rhlt77}.  \tcb{The theory predicts for oblate bodies a transition angle $\theta^*$ slightly greater than in the prolate case, but the difference is within one degree for $e\le 0.866$.}

\subsection{Analysis of three-dimensional dynamics near a vertical wall}
The fully three-dimensional equations, while more complicated, can still be investigated analytically. With $\phi\neq 0$ (and $\beta=0$ as before), the discussion in the preceding subsection remains relevant because $\dot{\theta}$ and $\dot h$ depend on $\phi$ only through the common factor $\cos\phi$ in Eqs.~\eqref{eq:hprimeV}-\eqref{eq:thetaprimeV}, and may be divided as before to make $h$ the independent variable\footnote{This division now implicitly assumes that $\cos\phi$ does not vanish on any open time interval, so we must note separately the steady \tcb{solutions at $\theta = 0$, $\phi=\pi/2$ (prolate) and $\theta=\pi/2$, $\phi = \pi/2$ (oblate).}}. The argument now proceeds in the same way and results in an incomplete but still valuable description of the three-dimensional orbit: the projection of the trajectory in $(h,\theta,\phi)$-space onto the $h\theta$-plane must lie on a single level set of $\Psi$ in \eqref{eq:Psi} as determined by the initial condition.  

\tcb{A difference between two- and three-dimensional trajectories that is visible in the $h\theta$-plane is that the projection of a three-dimensional trajectory may traverse only part of a contour of $\Psi$ instead of all of it. In particular, an orbit may be periodic even though the contour of $\Psi$ has an asymptote for some finite value of $\theta$.  This is possible because the periodic orbit repeatedly traverses a subset of the contour, doubling back on itself at regular intervals.  To explain this behavior, we note that $\cos\phi>0$ implies that a particle with $\theta<0$ is moving toward the wall, whereas for $\cos\phi<0$ the opposite holds.  The result is that initial conditions with $\phi=0$ for which \eqref{eq:LimitThetaPsi} predicts periodicity also lead to periodic trajectories when the initial $\phi$ is modified.}

\begin{figure}
 \[\includegraphics[width=\linewidth]{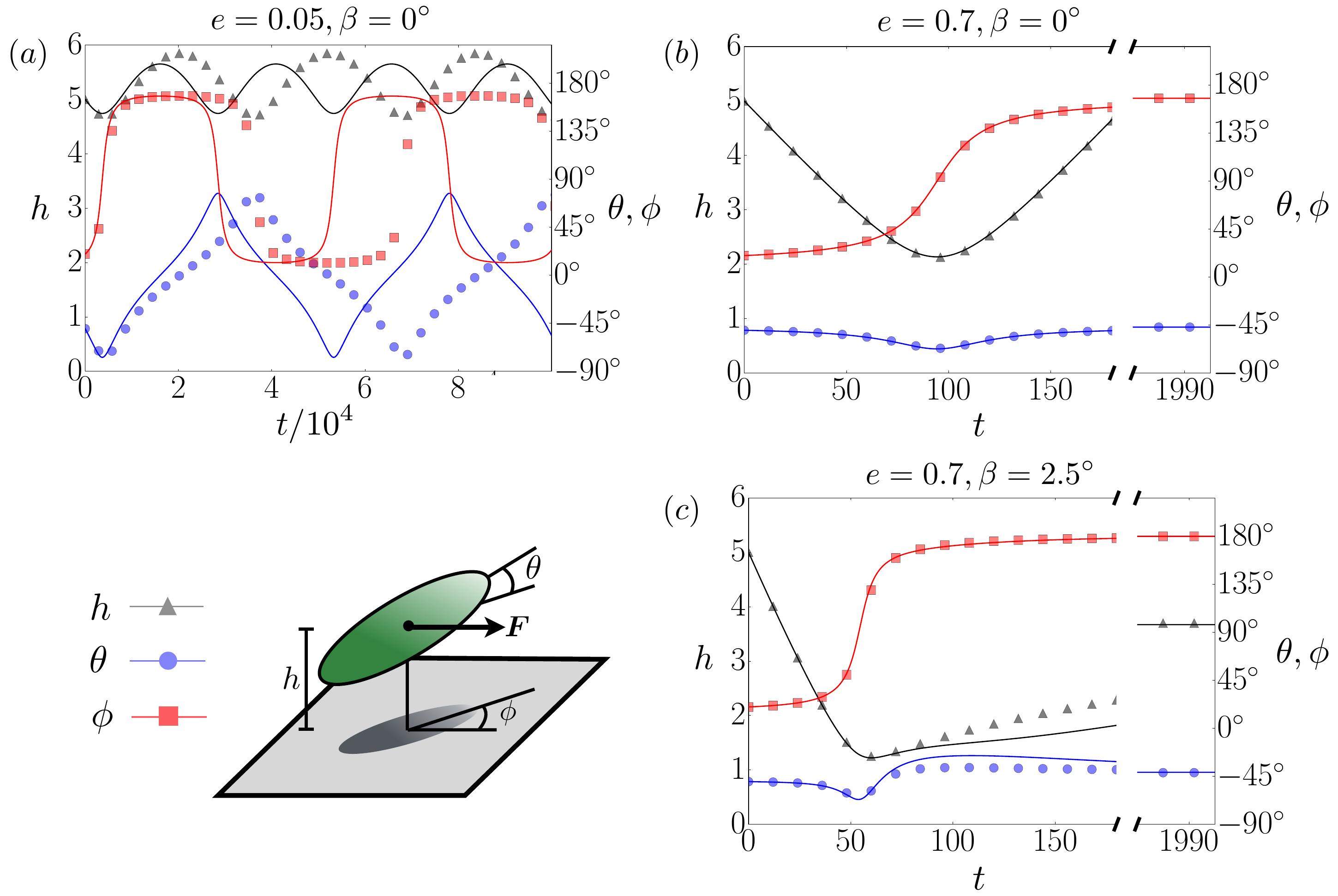}\]
\caption{Three-dimensional trajectories of sedimenting prolate spheroids, from full numerical simulations (symbols) and integration of the fourth-order accurate ODEs from \S\ref{sec:moi} (lines). (a) Three-dimensional periodic tumbling orbit of a nearly spherical particle near a vertical wall. The dynamics are akin to a three-dimensional reversing trajectory which fails to escape from the wall. The analytical prediction captures the shapes and amplitudes of the particle-wall distance $h$ and the orientation angles $\theta$ and $\phi$, but with an error in the frequency of oscillation. (b) A reversing trajectory of a more eccentric particle near a vertical wall. The body visits the wall and departs, settling to a constant final orientation in both $\theta$ and $\phi$ as $h\rightarrow \infty$. (c)  The same particle as in (b) near a slightly tilted wall converges to the stable sliding trajectory. The particle initially rotates while approaching the wall and then recedes towards a limiting separation distance and orientation, with $\phi\rightarrow \pi$ (the dynamics tend toward the two-dimensional sliding trajectory). }
\label{fig:lf}
\end{figure}

For a trajectory where $\phi\neq 0$, \eqref{eq:yprimeV} implies that $\dot{ y}\neq 0$, i.e. the particle will move laterally.  In the case of a periodic trajectory, the body wobbles periodically as it falls, drifting laterally back and forth along the wall \tcb{as described in \S\ref{sec:3Dgrw}}. \tcb{Plots of $h(t)$, $\theta(t)$, and $\phi(t)$} for such trajectories are shown in Fig~\ref{fig:lf}a, where we have set $e=.05$ (prolate), $\beta=0^\circ$, and initially $(h,\theta,\phi)=(5,-50^\circ,20^\circ)$. \tcb{An animation of such a trajectory is also provided as supplementary material}.  The dynamics are akin to a three-dimensional reversing trajectory which fails to escape from the wall. When the body is closest to the wall, $\theta$ is nearly $\pm \pi/2$, so that a small body rotation leads to a rapid change in $\phi$ from nearly zero to nearly $\pi$, or vice versa, and the body begins to drift laterally back in the direction from which it came. However, since the body is nearly spherical, the rotation induced by the wall is sufficient to rotate the major axis fast enough to redirect the body towards the wall yet again, and another reversing-type interaction ensues. 

Figure~\ref{fig:lf}b shows the trajectory for a more eccentric particle, with $e=0.7$, with the same initial condition, ($h,\theta,\phi)=(5,-50^\circ,20^\circ)$, which results in a complete three-dimensional reversing trajectory (also depicted in \tcb{Fig.~\ref{fig:AwesomeFigure}c)}. Just as in the previous case, when the body reaches the point nearest to the wall there is a rapid rotation in $\phi$, but in this case the body then ceases to rotate and escapes, settling to a constant orientation.  
\tcb{This limiting orientation has an interesting relationship to the dynamics near an inclined wall, to which we now turn}.

\subsection{Analysis of the fully general problem and the sliding trajectory}
We now consider the most general version of the problem, with an inclined wall ($\beta>0$) and fully three-dimensional sedimentation dynamics ($\phi\neq0$). A \tcb{non-zero} wall inclination angle reduces the symmetry in the problem and weakens the constraints of time-reversibility on the dynamics. \tcb{One consequence is that} there are no longer periodic orbits of the form discussed in the previous section. \tcb{Another is that} the three-dimensional dynamics can be driven towards the two-dimensional state, with $\phi$ reducing in magnitude to zero as $t\to \infty$. While the complete system presented in \tcb{Appendix~\ref{Appendix:general}} resists analytical treatment, the existence \tcb{and stability} of a sliding trajectory may still be investigated as follows.

Neglecting terms of $O\left(h^{-3}\right)$ in \eqref{eq:hprime}, \eqref{eq:thetaprime}, and \eqref{eq:phiprime} \tcb{in \tcb{Appendix~\ref{Appendix:general}}}, a fixed point (the {\it sliding} trajectory as depicted in Fig.~\ref{Trajboard}d) may be found explicitly. Taking $\phi=0$ (the two-dimensional, laterally symmetric case) gives $\dot\phi=0$, and then the reduced expression for $\dot\theta$ vanishes when $\theta=\theta_0$, where 
\begin{gather} \theta_0 = \frac{1}{2} \, \tan^{-1}\left(\frac{2}{3} \, \cot\left(\beta\right)\right),\end{gather}
an equation previously derived by \cite{hg94} using expressions from \cite{Wakiya59}.  
Finally with $\theta=\theta_0$ and $\phi=0$ the reduced expression for $\dot h =U_z$ vanishes when $h=h_0$, where 
\begin{gather}
\label{eq:h0_beta_e}h_0 = \frac{9 \, X^AY^A }{8 Y^A \pm 4{\left(X^A - Y^A\right)\left(\left(3+2\cot^2\beta\right)\left(9+4\cot^2\beta\right)^{-1/2}\pm1\right) }},
\end{gather}
\tcb{where the $\pm$ signs should be replaced by $+$ for the prolate case and by $-$ for the oblate case, and where $X^A$ and $Y^A$ also have different definitions as indicated in Table \ref{table:parameters}.} Linearizing the reduced system about $(h=h_0,\theta=\theta_0,\phi=0)$, this fixed point is found to be stable to arbitrary small perturbations for $\beta>0$ \tcb{(the eigenvalues associated with the linear system may be shown to always be negative)}. As an example of a body which is attracted to this stable sliding trajectory, we consider again a particle of eccentricity $e=0.7$, but near a wall tilted an an angle $\beta=2.5^\circ$. \tcb{Details for} the body trajectory are shown in Fig.~\ref{fig:lf}c, using once again the initial condition ($h,\theta,\phi)=(5,-50^\circ,20^\circ)$. Unlike in the vertical wall case, the body approaches the wall and rotates in a reversing-type manner, but then settles to a finite wall separation distance as $t\to \infty$. Meanwhile, the rotation in $\theta$ and $\phi$ (and the distance $h$) are no longer symmetric about the time at which the centroid is closest to the wall. Since $h$ remains bounded, the interaction with the wall continues to influence the rotational velocity of the body, and the body continues to rotate into the plane of the two-dimensional dynamics ($\phi\to\pi$).

\begin{figure}
\centering
\includegraphics[width=.55\textwidth]{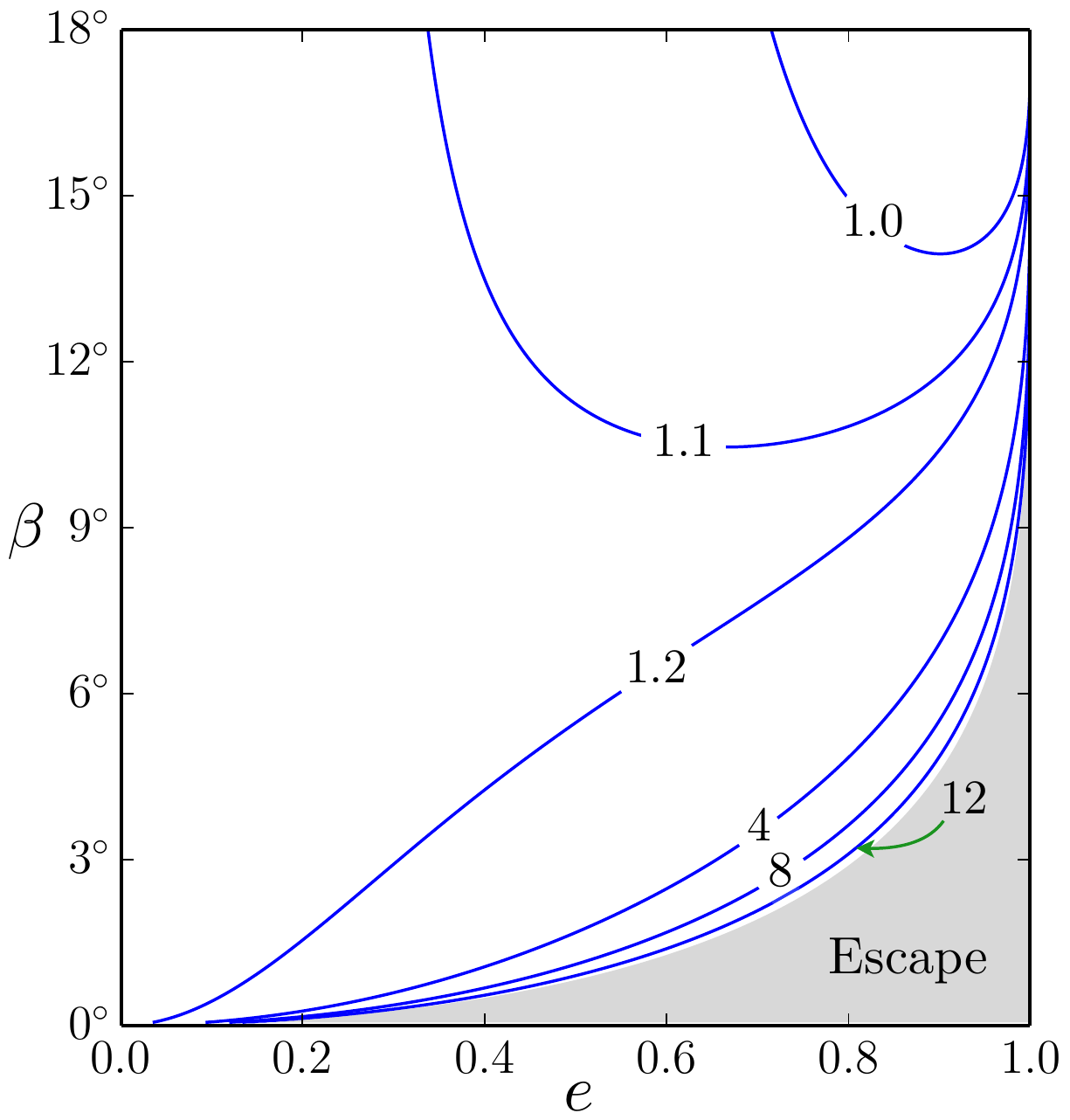} 
\caption{A prolate body near a sufficiently inclined wall cannot escape; it either approaches the wall so closely that particle or wall imperfections or other physics become important, or it assumes a stable orientation at a constant separation distance. In the latter case, the asymptotic separation $h_0$ is a function of the wall inclination angle and the particle eccentricity, with contours plotted above. Near the boundary of the escaping trajectories in the $e\beta$-plane one can find arbitrarily large values of $h_0$.}
\label{fig:H0vsBetaE}
\end{figure}

More generally, the equilibrium particle-wall separation $h_0$ is a function of the particle eccentricity and the wall inclination angle. The positive contours of $h_0$ for prolate bodies are plotted in the $e\beta$-plane in Fig.~\ref{fig:H0vsBetaE}.  Numerical simulations indicate that the contours with $h<2$ may overestimate the height of the fixed point, but the higher contours, near the boundary of the escaping trajectories, are reliable.

A sliding trajectory exists if the equilibrium particle-wall separation $h_0$ in \eqref{eq:h0_beta_e} is positive and finite. To ascertain whether an eccentricity $e$ and wall inclination angle $\beta$ result in a sliding trajectory, we set the denominator on the right-hand side of \eqref{eq:h0_beta_e} to zero. Solving for $\beta$ in terms of $e$ gives a critical wall inclination $\beta^*(e)$ beyond which the sliding trajectory arises. For $0<\beta<\beta^*(e)$, the wall is sufficiently vertical that the particle may escape if the initial condition is suitable. In this case $h_0$ is negative and the fixed point does not \tcb{describe} physical behavior.  The glancing and reversing trajectories are similar to their vertical-wall counterparts shown in Fig.~\ref{fig:AwesomeFigure}, except in that the orientation of the particle after the wall encounter need not be symmetric with its orientation beforehand. The rare cases of tumbling-type particles with positive $\beta$ exhibit perhaps the richest and most complex dynamics due to the low symmetry constraints.  These trajectories are not perfectly periodic; there is a gradual approach to the wall together with increased rotation rate until, possibly after many full revolutions, the particle collides with the wall.  

Meanwhile, if $\beta>\beta^*(e)$, escape is impossible and a particle beginning from any initial condition instead approaches the sliding fixed point whose coordinates are given (according to the $O(h^{-2})$ theory) above.  In a numerical study, \cite{kutteh2010} reported a second critical value of $\beta$ beyond which the sliding trajectory disappears and ``the particle monotonically approaches the wall until it makes contact.'' The analytical results presented here simply indicate a small particle-wall equilibrium distance (the small contour heights in Fig.~\ref{fig:H0vsBetaE}), but the underlying assumptions are not suitable for modeling close particle-wall interactions.  

The long formula for $\beta^*(e)$ (not shown here) reproduces the four values given in Table 9 of \citep{hg94} for prolate and oblate bodies of aspect ratios $c/a\in\{0.1,0.5\}$, providing a useful check on the method in a situation where the particle is far from the wall. In the limit of very slender prolate particles, as $e\rightarrow 1$, we find
\begin{gather}
\beta^*(e\rightarrow 1)=\cos^{-1}\sqrt{\frac{3}{11} \left(5-\sqrt{3}\right)}\approx19.25^\circ.
\end{gather}
For walls inclined more steeply than this angle, a prolate body of any eccentricity cannot escape. The drag anisotropy in the limit $e\to 1$ is not nearly as significant in the oblate case as it is in the prolate case (see \cite{hb55}), which implies that escape from the wall is more difficult; in fact a wall inclination greater than $11.48^\circ$ is sufficient to prevent escape for all oblate bodies.

\section{Discussion}
\label{sec:Discussion}
In this paper we studied the sedimentation of rigid prolate and oblate spheroids in a highly viscous fluid near a vertical or tilted wall. A system of ordinary differential equations governing the fully three-dimensional trajectories was derived. In numerous special cases, the system of equations yielded approximate analytical results for particle trajectories. The analytical predictions were compared to the results of full numerical simulations of the Stokes equations using a novel double layer boundary integral scheme, the method of stresslet images. These two approaches were used to investigate a wide array of trajectory types for bodies of arbitrary eccentricity, and near a vertical or inclined wall. When the wall is vertical, a nearly spherical body may undergo a periodic tumbling motion. \tcb{In three-dimensional versions of the tumbling trajectory, a drift-less periodic lateral wobbling arises.} For more eccentric particles, three-dimensional glancing and reversing trajectories appear, with the body approaching the wall only once before receding back into the bulk fluid. When the wall is tilted, the symmetry in the system is weakened. As a consequence, new trajectory types appear, while the periodic tumbling orbit vanishes. Glancing and reversing-type behavior is still possible, but a sliding trajectory emerges for many combinations of particle eccentricity and wall inclination angle. The sliding trajectory was found to be asymptotically stable to small translational and rotational perturbations in the far-field hydrodynamic theory. Critical wall inclination angles distinguishing sliding from either escaping or colliding with the wall were also presented.


Improvement of the analytical predictions given in this paper might be challenging. For instance, the inclusion of lubrication effects would be beneficial for understanding the near-wall interactions but would require other techniques similar to those employed for sphere-sphere interactions by \cite{dbb87} (see also \cite{bb88}). At the same time, the techniques we used here could be extended with no conceptual (but perhaps some algebraic) difficulty to deal with general triaxial ellipsoids.  The mobility problem for imposed torques can also be solved in a similar manner, which could be used to obtain the solution of the general resistance problem to the same level of accuracy. Body deformability, multiple-body interactions, and the inclusion of a background shear flow may be considered in future work.




\acknowledgements{We are grateful to Michael Graham, James Meiss, and Jean-Luc Thiffeault for helpful comments. We also acknowledge the authors of the freely available 3D and 2D visualization software packages Mayavi and Matplotlib \citep{ramachandran2011mayavi, Hunter:2007}. This research was supported in part by NSF grant DMS 1147523.}

\appendix
\section{Approximate sedimentation dynamics of spheroids in the general setting}\label{Appendix:general}
The solution of the mobility problem for a sedimenting prolate or oblate body near a wall with inclination angle $\beta$ is discussed in \S\ref{sec:moi}. The components of the translational velocity, $\b{U}=U_x \b{\hat{x}}+U_y\b{\hat{y}}+U_z\b{\hat{z}}$, are given by
\begin{align}
\nonumber U_x&= \frac{{\left( 2 \cos\beta-(1\pm\cos(2\theta)) \cos\beta\cos^2 \phi \pm \cos\phi\sin\beta \sin\left(2  \theta\right) \right)}
{\left(X^A - Y^A\right)}+ 2  Y^A \cos\beta}{2   X^AY^A } 
\\&\quad-\frac{9  \cos\beta}{16 h} \label{eq:xprime}
+ \frac{1}{128    h^{3} } \bigg[4   e^{2} \cos\phi\sin\beta \sin\left(2   \theta\right)
\\\nonumber&\quad+ {\Big(2  e^{2} {\left( \cos\left(2 \, \theta\right)\pm1\right)}
\cos^2\phi + 18 \, e^{2} \cos^2\theta -
{\left( 17\pm7\right)}e^{2}  + 16\Big)} \cos\beta\bigg], \\
\begin{split}
\label{eq:yprime}U_y&=
\frac{\sin\phi{\big(\pm\sin\beta \sin\left(2 \,
\theta\right)-{\left(1\pm \cos\left(2 \, \theta\right)\right)} \cos\beta
\cos\phi \big)}{\left(X^A-Y^A \right)}}
{2 X^A Y^A} 
\\&\quad
+\frac{e^{2}
\sin\phi{\big( 2 \sin\beta \sin\left(2 \, \theta\right)+{\left( \cos(2\theta)\pm1\right)} \cos\beta \cos\phi \big)} }{64 h^{3}}, 
\end{split}\\
\begin{split}
\label{eq:hprime}
U_z&= \frac{9   \sin\beta}{8   h } -
\frac{2   Y^A \sin\beta \pm {\left( \cos\beta \cos\phi
\sin\left(2   \theta\right) + (\cos(2\theta)\pm1)
\sin\beta\right)} {\left(X^A - Y^A\right)}
}{2 {X^A}Y^A } 
\\&\quad- \frac{e^{2}  \cos\beta
\cos\phi \sin\left(2   \theta\right) -    {\left(14  
e^{2} \sin^2\theta +(1\pm5)e^2-16 \right)}  \sin\beta}{32    h^{3}}. \end{split}
\end{align}

\begin{figure}
\centering
\includegraphics[width=.7\textwidth]{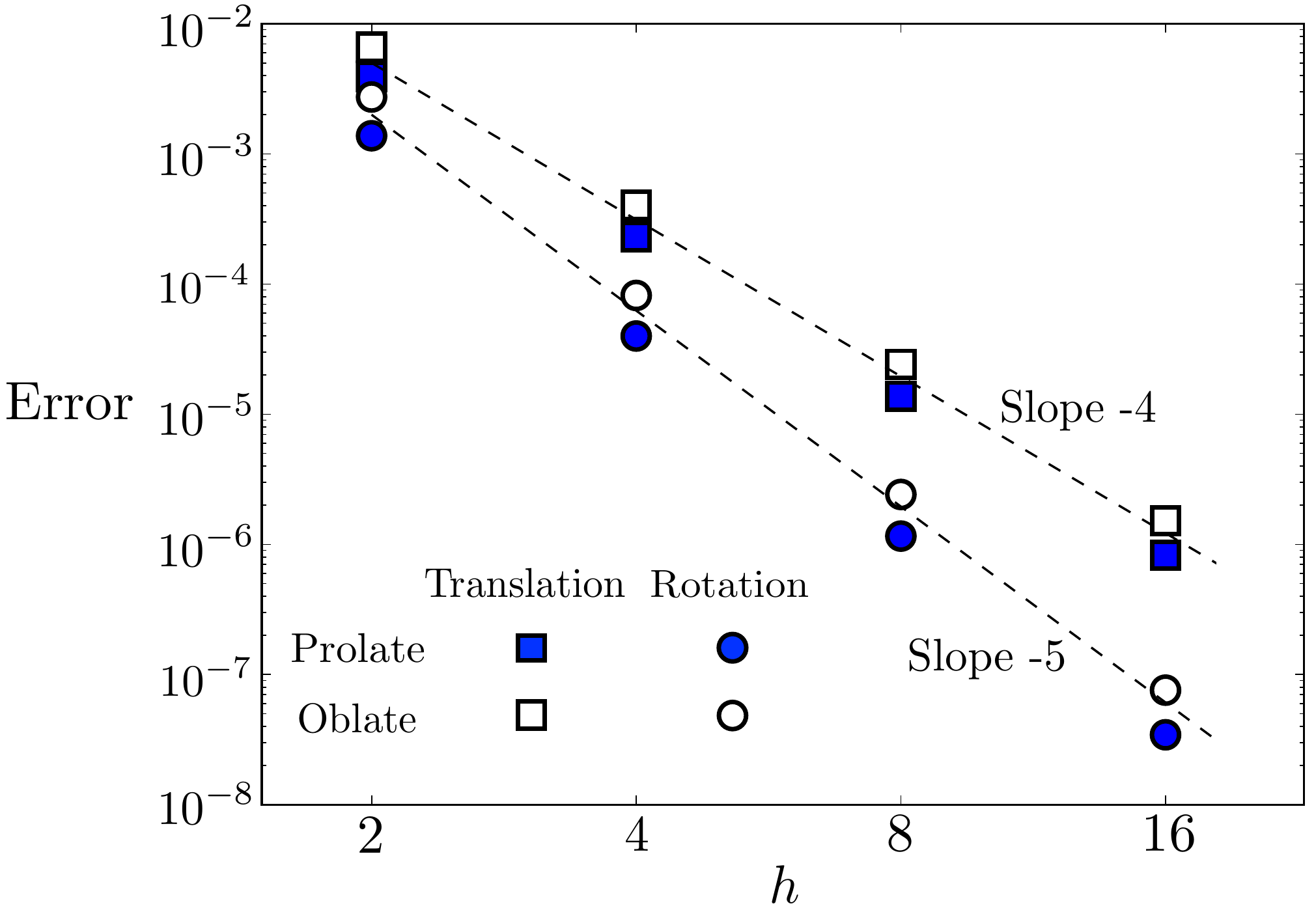}
\caption{\tcb{The boundary integral method and the differential equations agree with the expected rate of convergence in the distance from the wall: with errors scaling as $O(h^{-4})$ in the translational velocity and as $O(h^{-5})$ in the rotational velocity. Using $\theta=\pi/5$, $\phi = \pi/7$, $\beta = \pi/100$, and $e = \sqrt{3}/2$, the difference between the computed and analytical translational and rotational velocity vectors in the $\|\cdot \|_\infty$ norm are shown. Both prolate and oblate bodies are considered, using 1972 quadrature nodes for the prolate body and 2281 for the oblate body. This may be viewed either as a validation of the full numerical scheme, or as a check of the differential equations.}}
\label{fig:ConvergenceTest}
\end{figure}

The three components of the rotational velocity, $\bm{\Omega}=\Omega_x \b{\hat{x}}+\Omega_y\b{\hat{y}}+\Omega_z\b{\hat{z}}$, are given by 
\begin{align}
\label{eq:Omega-x}\begin{split} \Omega_x&=          	
\frac{9  e^{2}
\sin\phi {\left({\left(1\pm1-2   \sin^2\theta  \right)} \cos\beta \cos\phi - 3  
\sin\beta \sin\left(2   \theta\right)\right)} }{64   {\left(2-e^{2} \right)} h^{2}} 
\\
&\quad+ \frac{
6\cos\beta \cos\phi   {\left(6   e^{2} \sin^4\theta - 8\sin^2\theta
  + (1\pm1)(4-e^2-2e^{2}\sin^2\theta) \right)}}{128   {\left(e^{2} - 2\right)} h^{4}}\\
&\quad + \frac{3e^{2} \sin\beta \sin\left(2   \theta\right)\sin\phi{\left(12   e^{2}
\sin^2\theta + {\left(3\pm5\right)}e^{2}  -
18\right)} }{128   {\left(e^{2} - 2\right)} h^{4}},
\end{split}\\
\nonumber\Omega_y&= 
\frac{ 27e^2 \sin\beta\cos\phi \sin\left(2   \theta\right) 
+ 9 e^{2}\cos\beta{\left(2- 4   \cos^2\theta +  {\left(2   \cos^2\theta \pm1 - 1\right)}\sin^2\phi \right)}
 }{64   {\left(2-e^{2} \right)}
h^{2}}\\
\label{eq:Omega-y}\begin{split}
&\quad- \frac{3e^2\sin\beta\cos\phi \sin\left(2  \theta\right)}{128  {\left(2-e^{2} \right)} h^{4}}  
{\left(12   e^{2}\cos^2\theta - 5e^{2}(3 \pm1) + 18 \right)}  \\
&\quad-\frac{6\cos\beta}{128  {\left(2-e^{2} \right)} h^{4}} 
\Big[  
 e^{4} \cos^4\theta  {\left(6  \sin^2\phi - 9\right)} 
+ 2e^{2}\cos^2\theta \sin^2\phi   {\left(4 \pm e^{2} - 5e^{2} \right)} 
\end{split}\\
\nonumber&\quad+ e^{2}\sin^2\phi{\left(4-3  e^{2} \right)}  {\left(\pm1 - 1\right)}
- e^{2} \cos^2\theta{\left(12-(15\pm2)e^{2} \right)} 
- ( 7\pm1)e^{4}+ 10   e^{2} -4\Big],
\\ 
\label{eq:Omega-z}\Omega_z&= \cos\beta \sin\phi \sin\left(2\theta\right)\left[
 \frac{-9   e^{2} }{64   {\left(2-e^{2} \right)} h^{2}} 
 + \frac{3e^2   {\left(6   e^{2}\cos^2\theta - e^2(7\pm 5  ) + 8\right)} }{256   {\left(2-e^{2} \right)} h^{4}}\right].
\end{align}
The $\pm$ signs should be replaced with $+$ in the prolate case and $-$ in the oblate case.  The constants $X^A$ and $Y^A$ also have different definitions in these two cases, as indicated in Table \ref{table:parameters}. The errors in the expressions above are of size $\mathcal{O}(h^{-4})$ in the translational velocity and $\mathcal{O}(h^{-5})$ in the rotational velocity for $h \gg 1$ \tcb{(verified by comparison to full numerical simulations in Fig.~\ref{fig:ConvergenceTest} for a test problem with $\theta=\pi/5$, $\phi=\pi/7$, $\beta=\pi/100$, and $e = \sqrt{3}/2$)}. 

The time derivatives of $\phi$ and $\theta$ can be obtained from $\bm \Omega$ through the relations 
\begin{align}
\label{eq:dthetaFromOmega}\dot{\theta}&= -\Omega_y\cos(\phi) + \Omega_x\sin(\phi),\\
\label{eq:dphiFromOmega}\dot{\phi}&= 
\begin{cases}
\Omega_z - \tan(\theta)[\Omega_x\cos(\phi) + \Omega_y\sin(\phi)] & \textnormal{(prolate)},\\
\Omega_z + \cot(\theta)[\Omega_x\cos(\phi) + \Omega_y\sin(\phi)] & \textnormal{(oblate)}.
\end{cases}
\end{align}
\tcb{For degenerate geometries where $\phi$ is indeterminate ($\theta=0$ for oblate bodies and $\theta=\pi/2$ for prolate bodies) these formulas require the choice $\phi=0$; this degeneracy requires no extra bookkeeping in our problem since particles must already have $\phi=0$ when passing through these indeterminate positions.}
The equations \eqref{eq:dthetaFromOmega}-\eqref{eq:dphiFromOmega} can then be simplified to give
\begin{align}
\nonumber\dot{\theta}&=  \frac{18 e^2  \cos\beta \cos\phi
\cos\left(2   \theta\right) - 27e^2 \sin\beta \sin\left(2
  \theta\right)}{64   {\left(2 - e^{2} \right)}h^{2} }
\\&\quad+\frac{3 }{256 \, {\left(2-e^{2} \right)} h^{4}}\Big[4e^{2}\cos\theta\sin\beta \sin\theta
{\Big(18-{\left(9\pm5\right)}e^{2}  + 6 \,e^{2} \cos\left(2 \, \theta\right) \Big)} \label{eq:thetaprime}
\\\nonumber&\quad-\cos\beta\cos\phi{\Big(16- 16e^{2}+ 7 e^{4}
+e^2\cos\left(2 \theta\right)\left(24-{\left(12\pm4\right)}e^{2}+9 \, e^{2} \cos\left(2 \, \theta\right) \right)  
\Big)} \Big],\\
\label{eq:phiprime} \dot{\phi}&=
\begin{cases}
\displaystyle\frac{3\cos\beta\sin\phi\tan\theta}{64 {\left(2-e^{2} \right)}}
\left(\frac{-6  e^{2} }{ h^{2}} + \frac{3  e^{4}
\cos^2\theta - 8  e^{4} + 10  e^{2} - 4}{ h^{4} }\right) & \textnormal{(prolate)},\\
\displaystyle\frac{3\cos\beta\sin\phi\cot\theta}{64 {\left(2-e^{2} \right)}}
\left(\frac{-6  e^{2} }{h^{2}} - \frac{3  e^{4}\sin^2\theta - 2  e^{4} - 2  e^{2} - 4}{ h^{4} }\right)&\textnormal{(oblate)},
\end{cases}
\end{align}
\tcb{completing the} system of ODEs governing the evolution of $(x,y,h,\theta,\phi)$, as discussed in \S\ref{sec:moi}.

\section{Verification and validation of the numerical method}
In this section we give additional details on the accuracy of the method of stresslet images described in \S \ref{sec:Computation}.  This includes \tcb{a convergence study} and direct comparisons to previously published results for an inclined oblate body and exact solutions for the motion of a sphere.

\tcb{We begin by describing some checks on the time-stepping trajectory computations in this work. For the trajectories with lateral symmetry of Figs.~\ref{Trajboard} and \ref{fig:5traj}, the computed out-of-plane motion provides a simple quantitative estimate of the accumulated error; for these trajectories we rejected results for which the ratio of out-of-plane to in-plane translations or rotations exceeded $10^{-4}$ in any timestep.  
For the three-dimensional trajectories of Figs.~\ref{fig:AwesomeFigure} and \ref{fig:lf} this simple check is not available, so we reverse the direction of gravity after each computation and evolve back to the original position; in all of the results presented here the initial position was recovered with relative error less than $10^{-3}$ in the $\|\cdot\|_\infty$-norm. Typical trajectory runs used $\sim500$ nodes and $\sim250$ timesteps, i.e. required the inversion of 500 dense matrices of size 1500.  }  


\label{sec:convergence_tests}
\subsection{Convergence tests for an oblate body near a wall}
We now benchmark our numerical method against the results obtained by \cite{hg89}.  
For this test we consider an oblate body with unit radius and aspect ratio 2 or 10 (in our notation $e=\sqrt{3}/2$ or $e= 3\sqrt{11}/10$) and with position and orientation described by the parameters $h=1.1$ or $h=1.5$, $\theta = 75^\circ$, and $\phi=0$.  
We impose a wall-normal velocity $\bm U = (0,0,1)$ and zero rotation and solve the resistance problem, reporting the $z$-component of drag normalized by the value that would prevail in the absence of a wall.  These normalizations are known exactly, $15.084358$ for $a/c=2$ and $11.862466$ for $a/c=10$.  The results are indicated for various discretization levels $N$ in Table \ref{Table:HsuGanatosCheck} together with the values computed by \cite{hg89}.  The numerical method of that work was optimized to treat the case of an axisymmetric body and the quoted results were accordingly obtained more cheaply, with $N\le 100$.  In three of the four cases our results agree, but with $h=1.1$ and  $a/c=2$ there is a small difference which is likely due to insufficient grid resolution in the previously published work.

\begin{table}
 \centering \small
\begin{tabular}{ccccc}
\toprule 
\multirow{2}{*}{$N$}   & \multicolumn{2}{c}{ $h=1.5$} &   \multicolumn{2}{c}{ $h=1.1$} \\
  & $a/c=2$ &  $a/c=10$ & $a/c=2$ & $a/c=10$ \\
\midrule
36 & 2.370702 & 3.195427 & 5.003334 & 2.249052  \\
82 & 2.368974 & 1.857825 & 5.069687 & 2.773342  \\
160 &2.369387 & 1.839714 & 5.200719 & 2.774615 \\
331 &2.369652 & 1.838072 & 5.237369 & 2.733028\\
657 &2.369711 & 1.838784 & 5.240449 & 2.718526 \\
1286 & 2.369700 & 1.840077 & 5.240046 & 2.717739  \\
2484 & 2.369704 & 1.840311 & 5.240033 & 2.717475  \\
\midrule 
 \cite{hg89}  &  2.370 & 1.840 & 5.20 & 2.72 \\
\bottomrule
\end{tabular}
\caption{Convergence tests for a resistance problem with an inclined oblate spheroid of aspect ratio 2 or 10 and with centroid at height 1.1 or 1.5.  
with $N$ the number of nodes used to discretize the body surface. }
\label{Table:HsuGanatosCheck} \end{table}

\subsection{Comparison to exact solutions and regularized Stokeslets for a sphere near a wall}
\newcolumntype{g}{>{\columncolor{light-gray}}l}
\begin{table}
\centering \small
\begin{tabular}{l l rg rg rg rg c}
\toprule %
$\alpha$  & Gap Size & \multicolumn{8}{c}{Numerical $F^*$}  &Exact $F^*$\\
\cmidrule{3-10}
       &          &\multicolumn{2}{c}{$N=468$} & \multicolumn{2}{c}{$N=812$} 
 & \multicolumn{2}{c}{$N=1486$}  & \multicolumn{2}{c}{$N=2718$} &\\
\midrule
10 &  1101.22 &  0.9933 &  1.000061  &  0.9960 & 1.000048  &  0.9980 & 1.000057  &  0.9991 & 1.000050  &  1.000051\\ 
3.0 &  0.90677 &  1.0515 & 1.059071  &  1.0545 & 1.059058  &  1.0567 & 1.059067  &  1.0580 & 1.059060  &  1.059061\\ 
2.0 &  0.27622 &  1.1644 & 1.173821  &  1.1681 & 1.173807  &  1.1708 & 1.173818  &  1.1725 & 1.173809  &  1.173811\\ 
1.0 &  0.05431 &  1.5459 & 1.567463  &  1.5537 & 1.567449  &  1.5595 & 1.567467  &  1.5632 & 1.567451  &  1.567459\\ 
0.5 &  0.01276 &  2.0614 & 2.151222  &  2.0851 & 2.151566  &  2.1056 & 2.151459  &  2.1205 & 2.151471  &  2.151485\\ 
0.3 &  0.00453 &  2.4226 & 2.618852  &  2.4621 & 2.633134  &  2.5007 & 2.644156  &  2.5325 & 2.647208  &  2.647544\\ 
\bottomrule
\end{tabular}
\caption{Drag on a sphere of radius 0.1 translating parallel to a wall at six distances using four discretizations, normalized by the value predicted by Stokes' Law for an unbounded fluid.   
The shaded values were found using the method of stresslet images as described in \S \ref{sec:Computation}.  
These are accompanied by results obtained via regularized Stokeslets, quoted from Table 1 in \cite[]{adebc08}.}
\label{Table:ParallelSphere} 
\end{table}

\begin{table}
\centering \small
\begin{tabular}{l l rg rg rg rg r}
\toprule 
$\alpha$  & Gap Size & \multicolumn{8}{c}{Numerical $F^*$}  &Exact $F^*$\\
\cmidrule{3-10}
       &          &\multicolumn{2}{c}{$N=468$} & 
       \multicolumn{2}{c}{$N=812$}  & \multicolumn{2}{c}{$N=1486$}  & \multicolumn{2}{c}{$N=2718$} &  \\
\midrule
10 &  1101.22 &  1.0240  & 1.000104  & 1.0187 & 1.000087 & 1.0142 & 1.000101 & 1.0108 & 1.000104 & 1.000102    \\
3.0 &  0.90677 &  1.1556 & 1.125250  & 1.1488 & 1.125227 & 1.1431 & 1.125245 & 1.1388 & 1.125249 & 1.125246    \\
2.0 &  0.27622 &  1.4605 & 1.412894  & 1.4497 & 1.412842 & 1.4408 & 1.412873 & 1.4341 & 1.412878 & 1.412874    \\
1.0 &  0.05431 &  3.2441 & 3.036047  & 3.1952 & 3.035942 & 3.1557 & 3.036045 & 3.1266 & 3.036064 & 3.036064   \\
0.5 &  0.01276 &  10.287 & 9.216049  & 10.094 & 9.237569 & 9.9413 & 9.248925 & 9.8021 & 9.249230 & 9.251765  \\  
0.3 &  0.00453 &  19.707 & 19.43692  & 19.468 & 20.61629 & 20.247 & 22.54585 & 21.262 & 23.42283 & 23.66048 \\   
\bottomrule
\end{tabular}
\caption{Drag on a sphere of radius 0.1 translating perpendicular to a wall at six distances using four discretizations, normalized by the value predicted by Stokes' Law for an unbounded fluid.   
The shaded values were found using the method of stresslet images as described in \S \ref{sec:Computation}.  
These are accompanied by results obtained via regularized Stokeslets, quoted from Table 2 in \cite[]{adebc08}.}
\label{Table:NormalSphere} 
\end{table}

The resistance problem for a spherical body translating without rotation in a fluid bounded by a plane wall was solved exactly using bispherical coordinates by \cite{Oneill64} for motion parallel to the wall and by \cite{Brenner1961} for motion normal to it; see \cite{gcb67} for a summary of these results.  More recently, \cite{adebc08} solved this problem numerically using the method of regularized Stokeslets.  In this section we tabulate the results of the method of stresslet images against these exact and numerical solutions.  

The geometry of the problem is determined by two parameters, the sphere radius $a$ and the bispherical parameter $\alpha$ which satisfies $\cosh(\alpha)=d/a$, where $d$ is the distance from the particle center to the wall.  The gap size $d-a$ is the distance from the particle surface to the wall.  Following Ainley~{\it et al.}, we consider $a=0.1$ and $\alpha=10,3,2,1,0.5,0.3$.  The integrals over the sphere are discretized using $N$-point quadrature rules, for $N=468,812,1486,2718$.  We then calculate the drag $F$ and torque $T$ when the sphere translates at speed 1 with no rotation; these values are nondimensionalized by the drag and torque that would occur in the absence of the wall. 

In Table \ref{Table:ParallelSphere} we give the component of drag in the same direction as translation when the particle moves parallel to the wall, normalized by the drag predicted by Stokes' law for an unbounded fluid.   
Table \ref{Table:NormalSphere} gives this drag correction factor when the direction of translation is normal to the wall.  In both cases, the method of stresslet images gives more accurate results than the method of regularized Stokeslets at large and moderate gap sizes, and without the need of a regularization parameter; at the smallest gap size the performance of the two methods is similar. The exact solutions were recalculated following Equation (2.19) in the work of \cite{Brenner1961} and Equation (26) in the work of \cite{Oneill64}.  These values appear in the rightmost columns of Tables \ref{Table:ParallelSphere} and \ref{Table:NormalSphere}.  


\section{Initial data for model trajectories}\label{appendix:initialconditions}
For completeness, Table \ref{table:initialconditions} gives the initial data used as input to generate the model trajectories in Figures \ref{Trajboard} and \ref{fig:AwesomeFigure}.  
\begin{table}
\centering \small
\begin{tabular}{c c c c c c c c}
\toprule 
Figure and label & Shape & $e$ & $\beta$ & $h$ & $\theta$ & $\phi$ & Trajectory Type\\
\midrule
2a & Prolate & $0.980$  & $0$ & $3.0$ & $-20.00^\circ$ & $0$ & Glancing\\
2b & Prolate & $0.980$  & $0$ & $3.5$ & $-69.97^\circ$ & $0$  & Reversing\\
2c & Prolate & $0.150$  & $0$ & $3.0$ & $0$ & $0$& Periodic tumbling \\
2d & Prolate & $0.980$  & $9.17^\circ$ & $3.815$ & $37.53^\circ$ & $0$& Sliding \\
\midrule
3a & Prolate & $0.866$ & $0$ & $5.0$& $-34.38^\circ$&$-10.98^\circ$& 3D Glancing\\
3b & Oblate  & $0.866$ & $0$ & $9.69$& $-34.44^\circ$&$36.76^\circ$& 3D Glancing\\
3c & Prolate & $0.866$ & $0$ & $5.0$& $-60.00^\circ$&$-40.00^\circ$& 3D Reversing\\
3d & Oblate  & $0.866$ & $0$ & $5.0$ & $-60.00^\circ$ &$10.00^\circ$& 3D Reversing\\
\midrule
(Sup) & Prolate & $0.040$ & $0$ & $8.12$& $0.00^\circ$&$8.789^\circ$& 3D Periodic tumbling\\
(Sup) & Oblate  & $0.040$ & $0$ & $6.10$& $90.00^\circ$&$8.789^\circ$& 3D Periodic tumbling\\
\bottomrule
\end{tabular}
\caption{Parameters and initial conditions for the model trajectories depicted in Figures \ref{Trajboard} and \ref{fig:AwesomeFigure} as well for the prolate and oblate tumbling trajectories shown in the supplemental movie.  }  
\label{table:initialconditions} 
\end{table}

\bibliographystyle{jfm}
\bibliography{SedimentingEllipsoidsNotesREV}

\begin{thebibliography}{61}
\expandafter\ifx\csname natexlab\endcsname\relax\def\natexlab#1{#1}\fi

\bibitem[Ainley {\em et~al.\/}(2008)Ainley, Durkin, Embid, Boindala \&
  Cortez]{adebc08}
{\sc Ainley, J., Durkin, S., Embid, R., Boindala, P. \& Cortez, R.} 2008 The
  method of images for regularized {S}tokeslets. {\em J. Comput. Phys.\/} {\bf
  227}, 4600--4616.

\bibitem[Barta \& Liron(1988)]{bl88}
{\sc Barta, E. \& Liron, N.} 1988 Slender body interactions for low {R}eynolds
  numbers-part i: Body-wall interactions. {\em SIAM J. Appl. Math.\/} {\bf 48},
  992--1008.

\bibitem[Batchelor(1970)]{Batchelor70}
{\sc Batchelor, G.~K.} 1970 Slender-body theory for particles of arbitrary
  cross-section in {S}tokes flow. {\em J. Fluid Mech.\/} {\bf 44}~(03),
  419--440.

\bibitem[Batchelor(2000)]{Batchelor00}
{\sc Batchelor, G.~K.} 2000 {\em An introduction to fluid dynamics\/}.
  Cambridge University Press.

\bibitem[Blake(1971)]{Blake71}
{\sc Blake, J.~R.} 1971 A note on the image system for a {S}tokeslet in a
  no-slip boundary. In {\em Math. Proc. Cambridge\/}, , vol.~70, pp. 303--310.

\bibitem[Blake \& Chwang(1974)]{bc74}
{\sc Blake, J.~R. \& Chwang, A.~T.} 1974 Fundamental singularities of viscous
  flow. {\em J. Eng. Math.\/} {\bf 8}, 23--29.

\bibitem[Blake {\em et~al.\/}(2010)Blake, Tuck \& Wakeley]{Blake2010}
{\sc Blake, J.~R., Tuck, E.~O. \& Wakeley, P.~W.} 2010 A note on the
  s-transform and slender body theory in stokes flow. {\em IMA Journal of
  Applied Mathematics\/} {\bf 75}~(3), 343--355.

\bibitem[Bossis {\em et~al.\/}(1991)Bossis, Meunier \& Sherwood]{bms91}
{\sc Bossis, G., Meunier, A. \& Sherwood, J.~D.} 1991 Stokesian dynamics
  simulations of particle trajectories near a plane. {\em Phys. Fluids\/} {\bf
  3}, 1853--1858.

\bibitem[Brady \& Bossis(1988)]{bb88}
{\sc Brady, J.~F. \& Bossis, G.} 1988 Stokesian dynamics. {\em Annu. Rev. Fluid
  Mech.\/} {\bf 20}, 111--157.

\bibitem[Brenner(1961)]{Brenner1961}
{\sc Brenner, H.} 1961 The slow motion of a sphere through a viscous fluid
  towards a plane surface. {\em Chemical Engineering Science\/} {\bf
  16}~(3–4), 242--251.

\bibitem[Caro(2012)]{caro2012}
{\sc Caro, C.} 2012 {\em The Mechanics of the Circulation\/}. Cambridge
  University Press.

\bibitem[Chwang \& Wu(1975)]{cw75}
{\sc Chwang, A.~T. \& Wu, T.} 1975 {Hydromechanics of low-Reynolds-number flow.
  Part 2. Singularity method for Stokes flows}. {\em J. Fluid Mech.\/} {\bf
  67}, 787--815.

\bibitem[Chwang \& Wu(1976)]{cw76}
{\sc Chwang, A.~T. \& Wu, T.~Y.} 1976 Hydromechanics of low-{R}eynolds-number
  flow. {Part 4. T}ranslation of spheroids. {\em J. Fluid Mech.\/} {\bf 75},
  677--689.

\bibitem[Cichocki {\em et~al.\/}(2000)Cichocki, Jones, Kutteh \&
  Wajnryb]{cjkw00}
{\sc Cichocki, B., Jones, R.~B., Kutteh, R. \& Wajnryb, E.} 2000 Friction and
  mobility for colloidal spheres in {S}tokes flow near a boundary: The
  multipole method and applications. {\em J. Chem. Phys.\/} {\bf 112}~(5),
  2548--2561.

\bibitem[Cox(1970)]{Cox70}
{\sc Cox, R.~G.} 1970 The motion of long slender bodies in a viscous fluid part
  1. general theory. {\em J. Fluid Mech.\/} {\bf 44}, 791--810.

\bibitem[Durlofsky {\em et~al.\/}(1987)Durlofsky, Brady \& Bossis]{dbb87}
{\sc Durlofsky, L., Brady, J.~F. \& Bossis, G.} 1987 Dynamic simulation of
  hydrodynamically interacting particles. {\em J. Fluid Mech.\/} {\bf 180},
  21--49.

\bibitem[Edwardes(1892)]{Edwardes1892}
{\sc Edwardes, D.} 1892 Steady motion of a viscous liquid in which an ellipsoid
  is constrained to rotate about a principal axis. {\em Q. J. Math.\/} {\bf
  26}, 70--78.

\bibitem[Fax\'en(1922)]{Faxen1922}
{\sc Fax\'en, H.} 1922 {Der Widerstand gegen die Bewegung einer starren Kugel
  in einer z\"ahen Fl\"ussigkeit, die zwischen zwei parallelen Ebenen W\"anden
  eingeschlossen ist}. {\em Annalen der Physik\/} {\bf 4(68)}, 89--119.

\bibitem[Fax\'en(1924)]{Faxen1924}
{\sc Fax\'en, H.} 1924 {Der Widerstand gegen die Bewegung einer starren Kugel
  in einer z\"ahen Fl\"ussigkeit, die zwischen zwei parallelen Ebenen W\"anden
  eingeschlossen ist}. {\em Arkiv fur Matematik, Astronomi och Fysik\/} {\bf
  18(29)}, 1--52.

\bibitem[Goldman {\em et~al.\/}(1966)Goldman, Cox \& Brenner]{gcb66}
{\sc Goldman, A.~J., Cox, R.~G. \& Brenner, H.} 1966 The slow motion of two
  identical arbitrarily oriented spheres through a viscous fluid. {\em Chem.
  Eng. Sci.\/} {\bf 21}, 1151--1170.

\bibitem[Goldman {\em et~al.\/}(1967{\natexlab{{\em a\/}}})Goldman, Cox \&
  Brenner]{gcb67}
{\sc Goldman, A.~J., Cox, R.~G. \& Brenner, H.} 1967{\natexlab{{\em a\/}}}
  {Slow viscous motion of a sphere parallel to a plane wall -- I. Motion
  through a quiescent fluid}. {\em Chem. Eng. Sci.\/} {\bf 22}, 637--651.

\bibitem[Goldman {\em et~al.\/}(1967{\natexlab{{\em b\/}}})Goldman, Cox \&
  Brenner]{gcb67b}
{\sc Goldman, A.~J., Cox, R.~G. \& Brenner, H.} 1967{\natexlab{{\em b\/}}}
  {Slow viscous motion of a sphere parallel to a plane wall - II Couette flow}.
  {\em Chem. Eng. Sci.\/} {\bf 22}, 653--660.

\bibitem[Guazzelli(2006)]{Guazzelli06}
{\sc Guazzelli, {\'E}.} 2006 Sedimentation of small particles: how can such a
  simple problem be so difficult? {\em C. R. Mecanique\/} {\bf 334}, 539--544.

\bibitem[Happel \& Brenner(1983)]{hb55}
{\sc Happel, J. \& Brenner, H.} 1983 {\em {L}ow {R}eynolds number
  hydrodynamics: with special applications to particulate media\/}, , vol.~1.
  Springer.

\bibitem[Hsu \& Ganatos(1989)]{hg89}
{\sc Hsu, R. \& Ganatos, P.} 1989 The motion of a rigid body in viscous fluid
  bounded by a plane wall. {\em J. Fluid Mech.\/} {\bf 207}, 29--72.

\bibitem[Hsu \& Ganatos(1994)]{hg94}
{\sc Hsu, R. \& Ganatos, P.} 1994 Gravitational and zero-drag motion of a
  spheroid adjacent to an inclined plane at low reynolds number. {\em J. Fluid
  Mech.\/} {\bf 268}, 267--292.

\bibitem[Huang {\em et~al.\/}(1998)Huang, Hu \& Joseph]{hhj98}
{\sc Huang, P.~Y., Hu, H.~H. \& Joseph, D.~D.} 1998 Direct simulation of the
  sedimentation of elliptic particles in {O}ldroyd-b fluids. {\em J. Fluid
  Mech.\/} {\bf 362}, 297--326.

\bibitem[Hunter(2007)]{Hunter:2007}
{\sc Hunter, J.~D.} 2007 Matplotlib: A 2d graphics environment. {\em Computing
  In Science \& Engineering\/} {\bf 9}~(3), 90--95.

\bibitem[Jeffery(1922)]{Jeffery22}
{\sc Jeffery, G.~B.} 1922 The motion of ellipsoidal particles immersed in a
  viscous fluid. {\em Proc. R. Soc. A\/} pp. 161--179.

\bibitem[Johnson(1980)]{Johnson80}
{\sc Johnson, R.~E.} 1980 An improved slender-body theory for {S}tokes flow.
  {\em J. Fluid Mech.\/} {\bf 99}, 411--431.

\bibitem[Jung {\em et~al.\/}(2006)Jung, Spagnolie, Parikh, Shelley \&
  Tornberg]{jspst06}
{\sc Jung, S., Spagnolie, S.~E., Parikh, K., Shelley, M. \& Tornberg, A.-K.}
  2006 Periodic sedimentation in a {S}tokesian fluid. {\em Phys. Rev. E\/} {\bf
  74}, 035302.

\bibitem[Katz {\em et~al.\/}(1975)Katz, Blake \& Paveri-Fontana]{kbp75}
{\sc Katz, D.~F., Blake, J.~R. \& Paveri-Fontana, S.~L.} 1975 On the movement
  of slender bodies near plane boundaries at low {R}eynolds number. {\em J.
  Fluid Mech.\/} {\bf 72}, 529--540.

\bibitem[Keh \& Anderson(1985)]{ka85}
{\sc Keh, H.~J. \& Anderson, J.~L.} 1985 Boundary effects on electrophoretic
  motion of colloidal spheres. {\em J. Fluid Mech.\/} {\bf 153}, 417--439.

\bibitem[Keh \& Wan(2008)]{kw08}
{\sc Keh, H.~J. \& Wan, Y.~W.} 2008 Diffusiophoresis of a colloidal sphere in
  nonelectrolyte gradients perpendicular to two plane walls. {\em Chem. Eng.
  Sci.\/} {\bf 63}, 1612--1625.

\bibitem[Keller \& Rubinow(1976)]{kr76}
{\sc Keller, J.~B. \& Rubinow, S.~I.} 1976 Slender-body theory for slow viscous
  flow. {\em J. Fluid Mech.\/} {\bf 75}, 705--714.

\bibitem[Kim(1985)]{Kim85}
{\sc Kim, S.} 1985 Sedimentation of two arbitrarily oriented spheroids in a
  viscous fluid. {\em Int. J. Multiphase Flow\/} {\bf 11}, 699--712.

\bibitem[Kim(1986)]{Kim86}
{\sc Kim, S.} 1986 {Singularity solutions for ellipsoids in low-Reynolds-number
  flows: with applications to the calculation of hydrodynamic interactions in
  suspensions of ellipsoids}. {\em Int. J. Multiphase Flow\/} {\bf 12},
  469--491.

\bibitem[Kim \& Karrila(1991)]{kk91}
{\sc Kim, S. \& Karrila, S.~J.} 1991 {\em Microhydrodynamics: principles and
  selected applications\/}. Butterworth-Heinemann, Boston.

\bibitem[Kuiken(1996)]{kuiken96}
{\sc Kuiken, H.~K.} 1996 {H.A. Lorentz: Sketches of his work on slow viscous
  flow and some other areas in fluid mechanics and the background against which
  it arose}. {\em J. Eng. Math.\/} {\bf 30}, ii+1--18.

\bibitem[Kutteh(2010)]{kutteh2010}
{\sc Kutteh, R.} 2010 Rigid body dynamics approach to {S}tokesian dynamics
  simulations of nonspherical particles. {\em J. Chem. Phys.\/} {\bf 132}.

\bibitem[Lamb(1932)]{Lamb1932}
{\sc Lamb, H.} 1932 {\em Hydrodynamics\/}, 6th edn. Pergamon Press, New York.

\bibitem[Li {\em et~al.\/}(2013)Li, Manikantan, Saintillan \&
  Spagnolie]{lmss13}
{\sc Li, L., Manikantan, H., Saintillan, D. \& Spagnolie, S.~E.} 2013 The
  sedimentation of flexible filaments. {\em J. Fluid Mech.\/} {\bf 735},
  705--736.

\bibitem[Lighthill(1976)]{Lighthill76}
{\sc Lighthill, J.} 1976 Flagellar hydrodynamics. {\em Siam Rev.\/} {\bf 18},
  161--230.

\bibitem[Oberbeck(1876)]{Oberbeck1876}
{\sc Oberbeck, A.} 1876 {\"Uber station\"are Fl\"ussigkeitsbewegungen mit
  Ber\"ucksichtigung der inneren Reibung}. {\em J. {R}eine {A}ngew. {M}ath.\/}
  {\bf 81}, 62--80.

\bibitem[O'Neill(1964)]{Oneill64}
{\sc O'Neill, M.~E.} 1964 A slow motion of viscous liquid caused by a slowly
  moving solid sphere. {\em Mathematika\/} {\bf 11}, 67--74.

\bibitem[Power \& Miranda(1987)]{pm87}
{\sc Power, H. \& Miranda, G.} 1987 Second kind integral equation formulation
  of {S}tokes flows past a particle of arbitrary shape. {\em SIAM J. Appl.
  Math.\/} {\bf 47}, 689--698.

\bibitem[Pozrikidis(1992)]{Pozrikidis92}
{\sc Pozrikidis, C.} 1992 {\em Boundary integral and singularity methods for
  linearized viscous flow\/}. Cambridge University Press.

\bibitem[Pozrikidis(2007)]{Pozrikidis07}
{\sc Pozrikidis, C.} 2007 Particle motion near and inside an interface. {\em J.
  Fluid Mech.\/} {\bf 575}, 333--357.

\bibitem[Ramachandran \& Varoquaux(2011)]{ramachandran2011mayavi}
{\sc Ramachandran, P. \& Varoquaux, G.} 2011 {Mayavi: 3D Visualization of
  Scientific Data}. {\em Computing in Science \& Engineering\/} {\bf 13}~(2),
  40--51.

\bibitem[Russel {\em et~al.\/}(1977)Russel, Hinch, Leal \&
  Tieffenbruck]{rhlt77}
{\sc Russel, W.~B., Hinch, E.~J., Leal, L.~G. \& Tieffenbruck, G.} 1977 Rods
  falling near a vertical wall. {\em J. Fluid Mech.\/} {\bf 83}, 273--287.

\bibitem[Shapira \& Haber(1988)]{sh88}
{\sc Shapira, M. \& Haber, S.} 1988 {Low Reynolds number motion of a droplet
  between two parallel plates}. {\em Int. J. Multiphase Flow\/} {\bf 14},
  483--506.

\bibitem[Spagnolie \& Lauga(2012)]{sl12}
{\sc Spagnolie, S.~E. \& Lauga, E.} 2012 Hydrodynamics of self-propulsion near
  a boundary: predictions and accuracy of far-field approximations. {\em J.
  Fluid Mech.\/} {\bf 700}, 105--147.

\bibitem[Stakgold \& Holst(2011)]{sh11}
{\sc Stakgold, I. \& Holst, M.~J.} 2011 {\em Green's functions and boundary
  value problems\/}, , vol.~99. John Wiley \& Sons.

\bibitem[Steenberg \& Johansson(1958)]{sj58}
{\sc Steenberg, B. \& Johansson, B.} 1958 Viscous properties of pulp suspension
  at high shear-rates. {\em Svensk Papperstidning\/} {\bf 61}~(18), 696--700.

\bibitem[Stimson \& Jeffery(1926)]{sj1926}
{\sc Stimson, M. \& Jeffery, G.~B.} 1926 The motion of two spheres in a viscous
  fluid. {\em Proc. Roy. Soc. Lond. A\/} {\bf 111(757)}, 110--116.

\bibitem[Stokes(1851)]{Stokes1851}
{\sc Stokes, G.} 1851 On the effect of the internal friction of fluids on the
  motion of pendulums. {\em Trans. Camb. Phil. Soc.\/} {\bf 9}, 8.

\bibitem[Swaminathan {\em et~al.\/}(2006)Swaminathan, Mukundakrishnan \&
  Hu]{smh06}
{\sc Swaminathan, T.~N., Mukundakrishnan, K. \& Hu, H.~H.} 2006 Sedimentation
  of an ellipsoid inside an infinitely long tube at low and intermediate
  {R}eynolds numbers. {\em J. Fluid Mech.\/} {\bf 551}, 357--385.

\bibitem[Swan \& Brady(2007)]{sb07}
{\sc Swan, J.~W. \& Brady, J.~F.} 2007 Simulation of hydrodynamically
  interacting particles near a no-slip boundary. {\em Phys. Fluids\/} {\bf 19},
  113306.

\bibitem[Tillett(1970)]{tillett70}
{\sc Tillett, J. P.~K.} 1970 Axial and transverse {S}tokes flow past slender
  axisymmetric bodies. {\em J. Fluid Mech.\/} {\bf 44}, 401--417.

\bibitem[Wakiya(1959)]{Wakiya59}
{\sc Wakiya, S.} 1959 {Effect of a submerged object on a slow viscous flow
  (Report V). Spheroid at an arbitrary angle of attack}. {\em Res. Rep. Fac.
  Engng. Niigata Univ. (Japan)\/} {\bf 8}, 17--30.

\bibitem[Yang \& Leal(1983)]{yl83}
{\sc Yang, S.-M. \& Leal, L.~G.} 1983 Particle motion in {S}tokes flow near a
  plane fluid-fluid interface. {Part 1. S}lender body in a quiescent fluid.
  {\em J. Fluid Mech.\/} {\bf 136}, 393--421.

\end{thebibliography}

\end{document}